\documentclass[10pt]{iopart}
\usepackage{bm}
\usepackage[]{graphicx}
\usepackage{ulem}
\usepackage[tight]{units}
\usepackage{color}
\usepackage{nicefrac}
\usepackage{iopams}  
\usepackage[colorlinks=true,citecolor=blue,linkcolor= blue,linkcolor=blue,urlcolor=blue]{hyperref}

\newcommand{\cip}{$\sigma_{_{\|}}$}
\newcommand{\cpp}{$\sigma_{_{\perp}}$}
\newcommand{\ratio}[1]{\nicefrac{$\sigma_{_{\|}}$}{$\sigma_{_{\perp}}$}=#1}
\newcommand{\PFip}{$\text{PF}_{_{\|}}$}
\newcommand{\PFpp}{$\text{PF}_{_{\perp}}$}
\newcommand{\Sip}{$S_{_{\|}}$}
\newcommand{\Spp}{$S_{_{\perp}}$}

\newcommand{\Sipx}{$S_{_{\|}}$~}
\newcommand{\Sppx}{$S_{_{\perp}}$~}

\newcommand{\fs}[1]{Figs.~\ref{fig:#1}}

\newcommand{\Ff}[1]{figure~\ref{fig:#1}}
\newcommand{\Ffs}[1]{figures~\ref{fig:#1}}

\begin{document}
\newcommand{\text}[1]{\mathrm{#1}}

\title[]{Thermoelectric transport in strained Si and Si/Ge heterostructures}
\author{N. F. Hinsche$^1$\footnote{Corresponding author: nicki.hinsche@physik.uni-halle.de}, I. Mertig$^1$,$^2$ and P. Zahn$^1$,$^3$}
\address{$^1$ Institut f\"{u}r Physik, Martin-Luther-Universit\"{a}t Halle-Wittenberg, D-06099 Halle, Germany}
\address{$^2$ Max-Planck-Institut f\"{u}r Mikrostrukturphysik, Weinberg 2, D-06120 Halle, Germany}
\address{$^3$ Helmholtz-Zentrum Dresden-Rossendorf, P.O.Box 51 01 19, D-01314 Dresden, Germany}
\date{\today}

%

\begin{abstract}
 The anisotropic thermoelectric transport 
 properties of bulk silicon strained in [111]-direction were studied by detailed 
 first-principles calculations focussing on a possible 
 enhancement of the power factor. Electron as well as hole doping were examined 
 in a broad doping and temperature range. At low temperature and low doping 
 an enhancement of the power factor was obtained for compressive and tensile strain in the electron-doped 
 case and for compressive strain in the hole-doped case. For the thermoelectrically more important 
 high temperature and high doping regime a slight enhancement of the power factor was only found 
 under small compressive strain with the power factor overall being robust against applied strain. 
 To extend our findings the anisotropic thermoelectric transport of an [111]-oriented Si/Ge superlattice 
 was investigated. Here, the cross-plane power factor under hole-doping was drastically suppressed 
 due to quantum-well effects, while under electron-doping an enhanced power factor was found. With 
 that, we state a figure of merit of ZT$=0.2$ and ZT$=1.4$ at $T=\unit[300]{K}$ and $T=\unit[900]{K}$ 
 for the electron-doped [111]-oriented Si/Ge superlattice. All results are discussed in terms of band 
 structure features.
\end{abstract}

\submitto{Journal of Physics: Condensed Matter}

\maketitle


\section{Introduction}
Thermoelectric phenomena were first described for metals by Seebeck at the beginning of the 
19th century and revived by Ioffe in the late 1950's by the introduction of semiconductors to 
thermoelectric devices \cite{Seebeck,Ioffe1958}. 
However, since then thermoelectrics were restricted to a scientific and 
economic niche mainly due to their poor conversion 
efficiency \cite{Majumdar:2004p6568,Bottner:2006p2812}. 
Nowadays emerging global need for energy production and conservation 
has intensified interest and research in more effective alternative energy technologies 
to reduce our dependence on fossil fuels. 
Contributing to this, thermoelectric devices could partially convert wasted heat into electricity 
by their ability to transform heat directly into electric current, and vice versa \cite{Tritt:2006p15694}. 

The thermoelectric conversion efficiency can be stated
by the figure of merit (FOM)
\begin{equation}
ZT=\frac{\sigma S^{2}}{\kappa_{el} + \kappa_{ph}} T,
\label{eq1}
\end{equation}
where $\sigma$ is the electrical conductivity, $S$ the thermopower, $\kappa_{el}$  and 
$\kappa_{ph}$ are the electronic and phononic contribution to the thermal conductivity, respectively. 
The numerator of Eq.~\ref{eq1} is called power factor $\text{PF}=\sigma S^{2}$ and characterizes 
the electric power output.

While thermoelectric devices are extremely
facile, have no moving parts, and
do not produce greenhouse gases \cite{Sales:2002p6580}, two obstacles limit 
their applicability. The first, a low efficiency, could be challenged by 
nowadays nanostructured thermoelectrics enabling large values 
of $ZT \gg 1$ \cite{Venkatasubramanian:2001p114,Harman:2002p5345,Dresselhaus:2007p2775}.
As a second drawback, those materials are based on 
environmentally hazardous or rare lead, tellurium or selenium compounds and are therefore hard to integrate in 
semiconductor electronics.

However, current research gained a tremendous progress in enabling silicon for thermoelectrics. 
Silicon is non-toxic, readily available, cheap and well integrated in present electronics infrastructure, 
so this might be a considerable leap forward. While silicon has been 
stated as inefficient thermoelectric in the past due to its enormous thermal conductivity \cite{Vining:2008p9416}, 
recent experimental and theoretical attempts revealed that nanostructuring could lead to thermoelectric efficiencies comparable 
to state of the art commercial thermoelectric materials 
\cite{Hochbaum:2008p6569,Boukai:2008p14967,Bux:2009p14985,Tang:2010p15127}.

Besides the reduction of thermal conductivity, that is the denominator in Eq.~\ref{eq1}, Koga \etal~ showed in a seminal work \cite{Koga:1999p15445,Koga:1999p5363} 
that it should be possible to enhance the power factor, that is the numerator in Eq.~\ref{eq1}. 
This concept of carrier pocket engineering uses the influence of strain to optimize the band structure of 
silicon and germanium based superlattices (SL) regarding their electronic transport. 
As a main result it was found, that the effect of the lattice strain at the Si/Ge 
interfaces is more relevant for strain in [111], than in [001]-direction, regarding a possible enhancement of the power factor. 
In fact, $ZT = 1.25$ and $ZT = 0.98$ at room temperature were predicted for strain-non-symmetrized and strain-symmetrized 
[111]-oriented Si/Ge-SL, respectively, and the ZT values are shown to increase significantly at elevated temperatures \cite{Koga:2000p2542}.

While in a previous study we already concentrated on the influence of biaxial in-plane strain in [001]-direction 
on the thermoelectric properties of silicon \cite{Hinsche:2011p15276}, we will focus here on the 
influence of strain along the [111]-direction. 
For this purpose the paper is organized as follows. In section \ref{method} we introduce our approach footing 
on first principle 
electronic structure calculations within density functional theory and transport calculations 
based on the solution of the linearized Boltzmann equation. 
By this means we start the discussion of the thermoelectric 
transport properties of bulk silicon strained along the [111]-direction in dependence on 
strain and doping to gain inside into the physical mechanisms, which clearly differ from the 
[001]-strain case. The discussion is performed for electron- as well as for hole-doping in sec.~\ref{electron} 
and \ref{hole}, respectively. 

To extend the findings for the strained bulk silicon, in sec.~\ref{SiGe} results 
for an exemplary Si/Ge-SL grown on Si in [111]-direction are presented. 
Here the influence of tensile strain in [111]-direction, induced by the lattice mismatch at the Si/Ge 
interface, is investigated with respect to the thermoelectric transport in-plane and cross-plane the SL. 
Again, the temperature and doping dependence of the thermoelectric properties are discussed for electron- and hole-doping 
regarding a possible enhancement of the power factor. A further aspect will be the 
influence of structural relaxation and chemical composition on the transport properties. 
At the end of the paper, in sec.~\ref{FOM} insights into the FOM will be presented along with the electronic part 
of the thermal conductivity, to give a clue on optimal charge carrier concentrations to obtain the best FOM.

While focussing our interest in the high-temperature 
thermoelectric application of strained silicon, our results in the 
room temperature regime could be of importance 
for the metal-oxide-semiconductor device community. Knowledge of the thermoelectric properties 
of silicon under strain could help to understand parasitic effects on the electronic transport in those structures. 
In this low temperature and low doping regime 
we confirm a remarkable influence of externally applied strain on the electrical transport under electron and hole doping.  


\section{\label{method} Methodology}
Our approach is based on two constituents: first principles density functional theory calculations (DFT), 
as implemented in the \textsc{QuantumEspresso} package \cite{Giannozzi:2009p14969} and an in-house developed Boltzmann 
transport code \cite{Hinsche:2011p15276,Mertig:1999p12776,Hinsche:2011p15707,Zahn:2011p15523}  to calculate 
the thermoelectric properties.

First, the band structure of the strained and unstrained Si was calculated using 
the general gradient approximation (GGA) with the Perdew-Burke-Ernzerhof (PBE) flavor of exchange correlation functional  
\cite{Perdew:1996p14792}. Fully relativistic and norm-conserving pseudo potentials \cite{Corso:2005p8612} were 
used to treat the spin-orbit splitting of the Si valence bands appropriately. The calculations for the bulk Si 
were performed with the rhombohedral experimental 
lattice constant $a_0=5.434${\AA}/$\sqrt{2}$ for a rhombohedral two atom unit cell, which is sketched in the inset 
of \Ff{1}(c). 
The strain in [111]-direction under constant volume is simulated by changing the lattice constant a and with that 
angle $\alpha_r$. 
Throughout the paper the biaxial strain will be given in units of the relative change of the in-plane
lattice constant, that is the nearest neighbour distance in the [111]-plane as $\Delta a = a_{[111]}$ 
following the notation
of previous works~\cite{Hinsche:2011p15276,Hinsche:2011p15707,Yu:2008p14181,Bouhassoune:2009p6886}. 
The angle $\alpha_r$ is given by $\cos \alpha_r = 1-\frac{3(1+\Delta a/a_0)^6}{4+2(1+\Delta a/a_0)^6}$.
That means, tensile in-plane strain considers changes $\nicefrac{\Delta a}{a_{0}}>0$ and 
$\alpha_r >60^\circ$, while compressive in-plane strain 
means $\nicefrac{\Delta a}{a_{0}}<0$ and $\alpha_r <60^\circ$. 
As used previously in literature, tensile strain along [111]-direction coincides with 
compressive in-plane strain as denoted here.

By this trigonal deformation an atomic relaxation of the atomic positions inside the unit cell is possible, 
as the displacement of the two sublattices along [111]-direction is no longer given by symmetry. 
To obtain the atomic positions of the strained silicon we performed structural relaxations 
using \textsc{VASP}~\cite{Kresse:1996p12346}. The atomic positions were optimized such that the 
Hellmann-Feynman forces on them were below $\unit[0.1]{meV/\AA}$. At the same time 
the given deformed lattice parameters were not allowed to relax and conservation of the unit cell volume 
was assumed. A volume relaxation at the maximum strain $\nicefrac{\Delta a}{a_{0}} \pm 1\%$ 
resulted in a volume reduction by $0.2\%$ which corresponds to a lattice constant change by less than 
$0.1\%$.


As expected, our DFT calculations underestimate the size of the band gap at zero temperature and do not reproduce the 
temperature dependence of the gap. 
For this purpose we implemented a temperature-dependent scissor operator~\cite{Godby:1988p14795}, so that
the strain- and temperature-dependent energy gap $E_g$ becomes 

\begin{eqnarray}
\label{teg}
E_{g}(T,\frac{\Delta a}{a_0}) &=& E_{g}(T=0,\frac{\Delta a}{a_0}) + U_{\text{GGA}} \\ \nonumber
&-& \frac{ \alpha T^{2}}{T+\beta},
\end{eqnarray}

where $E_{g}(T=0,\frac{\Delta a}{a_0})$ is the zero temperature gap obtained by our self-consistent DFT calculations, $U_{\text{GGA}}=\unit[0.57]{eV}$ 
is a static correction to reproduce the experimental low temperature gap and the third part of Eq.~\ref{teg} is the correction of the 
temperature dependence of the band gap in a wide temperature range \cite{Varshni:1967p14976}, 
with $\alpha=\unit[4.73 \times 10^{-4}]{\nicefrac{eV}{K}}$, T the absolute temperature 
and $\beta=\unit[636]{K}$ for bulk silicon. 


Converged results from the first step are basis to obtain the thermoelectric transport properties 
by solving the linearized Boltzmann equation in relaxation time approximation (RTA) \cite{Mertig:1999p12776}. 
Boltzmann transport calculations for thermoelectrics have been carried out for quite a long time and show 
reliable results for metals~\cite{Vojta:1992p1395,Thonhauser:2004p14960} as well as for wide- and narrow-gap semiconductors~\cite{Hinsche:2011p15276,Hinsche:2011p15707,Singh:2010p14285,Scheidemantel:2003p14961} 
in the diffusive limit of transport.
Here the relaxation time is assumed to be constant with respect to wave vector k and energy on the scale of $k_{B}T$. 
The constant relaxation time allows for the calculation of the thermopower $S$ without any free parameter.
To reproduce experimental findings we parametrized doping dependent relaxation times 
from mobility measurements on unstrained silicon according to ref.~\cite{Jacoboni:1977p14945} by 
\begin{eqnarray}
\label{tau}
\tau(N)=\big( (&-&\nicefrac{c}{\pi} \cdot \arctan [a \cdot \lg(\nicefrac{N}{N_0})] + \nicefrac{c}{2}) \nonumber \\
&+& 2 (\lg (\nicefrac{N}{N_1})^{2} \big) \cdot d
,\end{eqnarray}
with $a=1.8, N_0=\unit[10^{17}]{cm^{-3}}, N_1=\unit[10^{17.5}]{cm^{-3}}, c=1500, d=\unit[0.15]{fs}$ ($a=1.3, N_0=\unit[10^{16.8}]{cm^{-3}}, N_1=\unit[10^{17.5}]{cm^{-3}}, c=550, d=\unit[0.13]{fs}$) for electron (hole) doping 
and charge carrier concentrations of $N$ between $\unit[10^{14}]{cm^{-3}}$ and $\unit[10^{22}]{cm^{-3}}$. 
Nevertheless, we state that our relaxation time is not strain-dependent, while 
it is known, that under strain the dominant scattering
process alters: For unstrained Si, the room-temperature scattering
is dominated by optical phonons, i.e., intervalley scattering,
whereas for strained Si, the scattering by optical
phonons is reduced \cite{Dziekan:2007p1770,Roldan:2012p15129}.

With the transport distribution function (TDF) as termed by Mahan and Sofo \cite{Mahan:1996p508}
\begin{eqnarray}
&\mathcal{L}_{\perp, \|}^{(n)}(\mu, T)= \nonumber \\
&\frac{\tau}{(2\pi)^3} \sum \limits_{\nu} \int d^3\mathbf{k} \left( v^{\nu}_{\mathbf{k},(\perp, \|)}\right)^2 (E^{\nu}_{\mathbf{k}}-\mu)^{n}\left( -\frac{\partial f_{(\mu,T)}}{\partial E} \right)_{E=E^{\nu}_\mathbf{k}} \nonumber
\\
\label{Tcoeff}
\end{eqnarray}
the temperature- and doping-dependent electrical conductivity $\sigma$ and thermopower $S$ are defined as
\begin{eqnarray}
\sigma_{_{\perp, \|}}=e^2 \mathcal{L}_{\perp, \|}^{(0)}(\mu, T) \qquad 
S_{_{\perp, \|}}=\frac{1} {eT} \frac{\mathcal{L}_{\perp, \|}^{(1)}(\mu,T)} {\mathcal{L}_{\perp, \|}^{(0)}(\mu,T)}
.\label{Seeb}
\end{eqnarray}
Here $\|$ denotes the in-plane direction parallel to the a-axis and $\perp$ the 
cross-plane direction. 
The electronic part to the total thermal conductivity accounts to
\begin{equation}
\kappa_{el}{_{\perp, \|}}=\frac{1}{T} \left(\mathcal{L}_{\perp, \|}^{(2)}(\mu,T)-\frac{(\mathcal{L}_{\perp, \|}^{(1)}(\mu,T))^2}{\mathcal{L}_{\perp, \|}^{(0)}(\mu,T)} \right) \, .
\label{kel}
\end{equation}
$E_\mathbf{k}^{\nu}$ denotes the band structure of band $\nu$, $v_\mathbf{k}^\nu$ the group velocity and $f_{(\mu,T)}$ 
the \textsc{Fermi-Dirac}-distribution with chemical potential $\mu$. 
The chemical potential $\mu$ at temperature $T$ and extrinsic carrier concentration $N$ is 
determined by an integration 
over the density of states $n(E)$
\begin{eqnarray}
N=\int \limits_{\mu-\Delta E}^{\text{VBM}} \text{d}E \,  n(E) [f_{(\mu,T)}-1]+
\int \limits_{\text{CBM}}^{\mu+\Delta E} \text{d}E \, n(E) f_{(\mu,T)}
\label{Dop},
\end{eqnarray} 
where $\text{CBM}$ is the conduction band minimum and $\text{VBM}$ is the 
valence band maximum. The necessary size of $\Delta E$ will be discussed below. 

In a recent work we showed, that the determination of 
surface integrals in anisotropic Brillouin zones is demanding with respect to 
convergence of the transport property anisotropy ~\cite{Zahn:2011p15523}. 
So, the constant energy-surface integrations, which are required in equation \ref{Tcoeff}, are performed within 
an extended tetrahedron method \cite{Lehmann:1972p14972,Zahn:1995p14971,Mertig:1987p5922} interpolating the 
calculated Eigenvalues $E_\mathbf{k}^{\nu}$ on an adaptive $\mathbf{k}$-mesh 
corresponding to a density of at least 44000 $\mathbf{k}$ points in the irreducible 
part of the Brillouin zone. $\mathcal{L}_{\perp, \|}^{(0)}(E, T=0)$ was determined 
on a dense energy mesh with a step width of $\unit[1]{meV}$. At vanishing strain $\nicefrac{\Delta a}{a_{0}}=0$ 
the numerical errors of \cip/\cpp and \Sip/\Spp were constantly below $0.1\%$.
In the limit of low carrier concentrations $N \leq \unit[1\times 10^{14}]{cm^{-3}}$ and for larger 
carrier concentrations in the bipolar conduction regimes at high temperatures, 
convergence of the integrals \ref{Tcoeff} and \ref{Dop} was achieved with an adaptive integration method 
for $\Delta E$ of at least $10k_b T$.

\section{Thermoelectric transport}
\subsection{\label{electron} [111]-strained silicon: electron doping}

\begin{figure}[h!]
\centering
\includegraphics[width=0.58\textwidth]{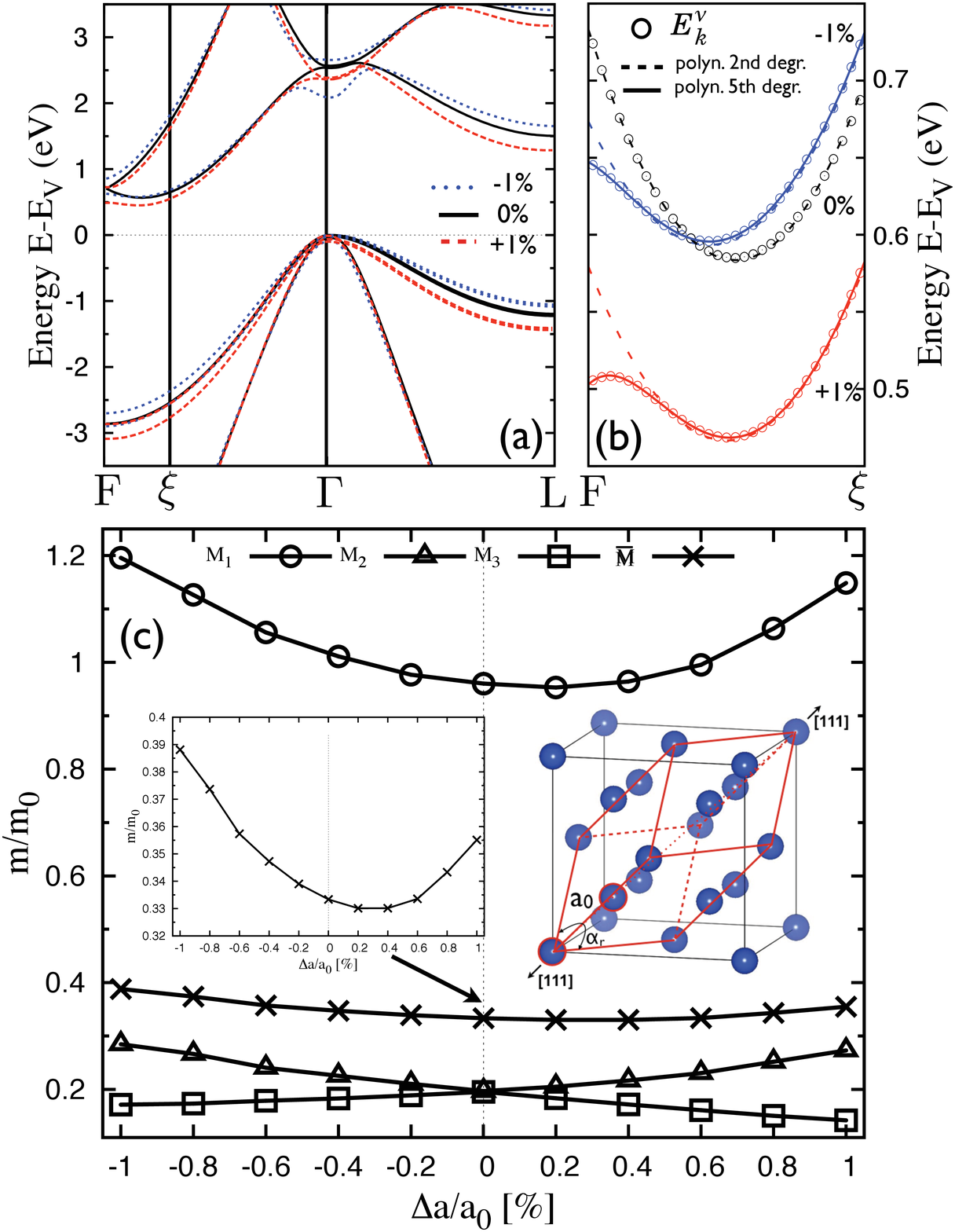}
\caption{\label{fig:1}(color online) Band structure of bulk silicon in the rhombohedral unit cell. In the unstrained case, 
the F point coincides with the X point of the fcc Brillouin zone. (a) The bands for the unstrained case (black solid line), 
under 1\% compressive strain (blue dotted line) and under 1\% tensile strain (red dashed line) are shown. In (b) a zoom near the conduction band 
minimum elucidates the nonparabolicity of the bands under applied strain. (c) Directional effective conduction band masses for silicon under 
[111]-strain. The insets show in detail the behaviour of the averaged effective mass 
$\overline{\text{M}}$ on the applied strain, as well as the rhombohedral unit cell~\cite{Momma} (red lines). More details are given in the text.}
\end{figure}

In \Ff{1}(a) the uncorrected band structure of bulk silicon in the rhombohedral unit cell is shown for the unstrained case 
(black solid line), under $1\%$ tensile strain (red dashed line) and for $-1\%$ compressive strain (blue dotted line) on relevant 
high symmetry lines. Unstrained silicon has an indirect band gap with conduction band minima (CBM)
near the $F$ high symmetry point. The CBM consists of sixfold degenerate ($\Delta_6$) prolate spheroidal isoenergetic 
surfaces along six equivalent $\Gamma -F$ directions. Due to symmetry of the lattice distortion in [111]-direction 
this degeneracy holds under applied strain in contrast to strain applied along [001]-direction \cite{Hinsche:2011p15276}. 

In the unstrained case, for each $\Delta_6$-valley the effective 
masses along the major and the minor axis are $\text{M}_1=0.91 m_0$ and $\text{M}_2=0.19 m_0$, respectively. 
As can be seen 
from \Ff{1}(b) the idea of an effective mass determined by a second order polynomial fit (dashed lines in \Ff{1}(b)) is valid 
for the unstrained case, but band warping leads to deviations for silicon under strain already for small band occupations. 
Here, a fifth order polynomial fit (solid lines in \Ff{1}(b)) is necessary to reproduce the band dispersion, 
which occurs under applied biaxial [111]-strain. It is therefore advisable to go beyond a 
simplified effective mass model. 
Contrary to the conduction bands, the [111]-strain leads to a splitting of the degenerate valence bands, 
the heavy-hole (HH) and light-hole band (LH), similar to the 
case of [001]-strain \cite{Hinsche:2011p15276,Baykan:2012p14974}. While the spin-orbit-split-off band 
is \unit[40]{meV} away from the band edge, the HH band lifts up energetically under tensile strain, while the LH 
band lowers in energy. This picture reverses under opposite strain conditions \cite{Sun:2007p14975}. 
While the indirect gap closes linearly under 
tensile in-plane strain from $\unit[0.58]{eV}$ in equilibrium to about $\unit[0.45]{eV}$ at $\nicefrac{\Delta a}{a_{0}}=1\%$, 
the gap size is almost constant within $\unit[0.01]{eV}$ under compressive strain within the considered range~\cite{Bouhassoune:2009p6886,Boykin:2007p15488,Niquet:2009p15435} (comp. \Ff{1}(b)). 
The direct gap at  $\Gamma$ decreases slightly under applied strain, more pronounced under compressive in-plane strain.

The influence of biaxial in-plane strain on the fitted effective masses is summarized in \Ff{1}(c). The change 
of the effective mass M$_1$ along the major axis is almost symmetric to the applied 
strain and increases up to $130\%$ of the unstrained value. In contrast, 
the transverse effective masses behave drastically different. While the transverse effective mass M$_2$ increases under 
tensile strain, the effective mass M$_3$ perpendicular to M$_2$ decreases. This behaviour is reversed under 
applied compressive strain. These results are in good agreement to other findings \cite{Boykin:2007p15488,Niquet:2009p15435}, 
but add up to GW calculations were only one transverse mass with an 
almost constant value was found~\cite{Bouhassoune:2009p6886}. 
The strain-dependent averaged effective 
mass $\overline{\text{M}}=\prod \limits_{i=1}^{3}(\text{M}_i)^{1/3}$, often referred to as density-of-states effective mass, 
is shown additionally as an inset in \Ff{1}(c).

\begin{figure*}[t]
\centering
\includegraphics[width=0.80\textwidth]{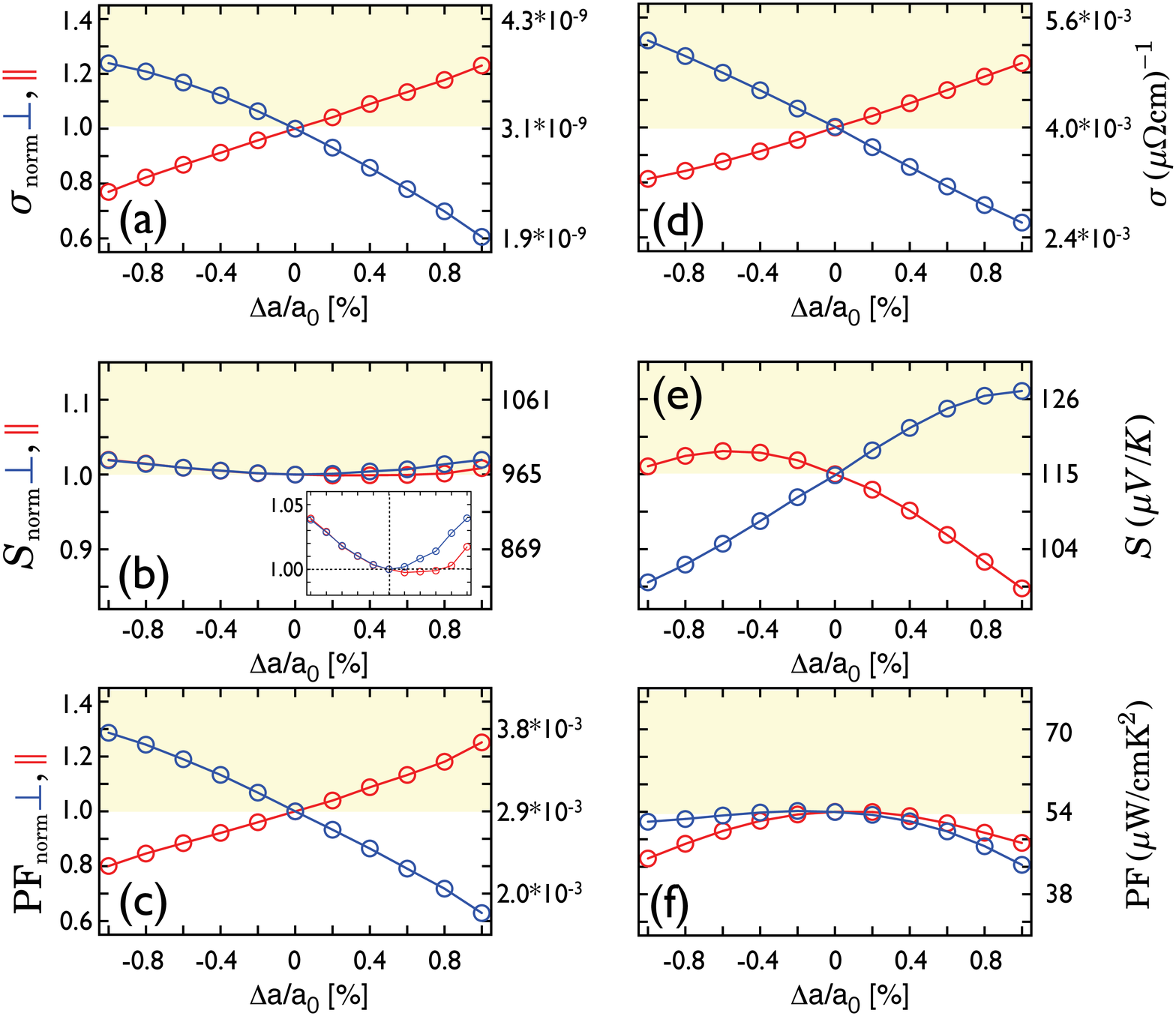}
\caption{\label{fig:2}(color online) Anisotropic thermoelectric transport properties of Si for fixed temperature 
and electron doping concentrations in dependence on compressive and tensile strain in [111]-direction. Left panels (a)-(c)) 
correspond to an electron doping of $\unit[2\times 10^{-8}]{e/atom}$ ($N = \unit[1\times 10^{15}]{cm^{-3}}$)
at a temperature of 100~K, while right panels (d)-(f) refer to an electron doping of $\unit[0.04]{e/atom}$ ($N = \unit[2\times 10^{21}]{cm^{-3}}$)
at a temperature of 900~K. On the left axis of each figure the relative value compared to the 
unstrained case is shown, while on the right axis the absolute values are given.}
\end{figure*}

In the following the influence of trigonal distortion on the thermoelectric transport of Si under 
electron doping will be discussed. For this purpose two doping and temperature regimes are considered. 
The first, at a low temperature of $T=\unit[100]{K}$ and low charge carrier concentration of 
$N = \unit[1\times 10^{15}]{cm^{-3}}$, is suitable for metal-oxide-semiconductor device applications. 
The results for the electrical conductivity, thermopower and power factor in dependence on the in-plane 
strain are shown in \Ff{2}(a),(b) and (c), respectively. Under tensile strain the in-plane electrical conductivity 
\cip increases almost linearly, while the cross-plane component \cpp decreases almost comparable. 
For compressive strain the behaviour reverses, with the cross-plane conductivity being enhanced up to 
23 \% at 1\% compressive strain, while the in-plane conductivity decreases to nearly 23\% of the unstrained value 
under 1\% tensile strain. 
In the limit of a degenerate semiconductor at low temperatures 
and small charge carrier concentrations this results can be completely understood within an effective mass 
calculation \cite{Zahn:2011p15523}. With noticeable variation of the electrical conductivity under applied strain, 
the thermopower is almost unaffected. At low temperatures only a small energy window near the band edges 
plays an important role for the determination of the thermopower. As the functional change of the 
coefficients $\mathcal{L}_{\perp, \|}^{(0,1)}(\mu, T)$ is determined by parabolic bands and a strain-dependent shift of 
the chemical potential, a strong change in the thermopower 
can not be expected. However, a slight upward tendency for the thermopower under compressive and 
tensile strain can be stated. As $S \propto (E_{\text{CBM/VBM}}-\mu)^{-1} \propto \overline{\text{M}}$, this can be directly 
linked to the strain-dependence of $\overline{\text{M}}$ (cf. \Ff{1}(c)). For tensile strain an anisotropy of 
the thermopower is apparent, which seems to be suppressed for compressive strain. 
This could be linked to stronger deviations from the isotropic effective mass for tensile strain as shown in \Ff{1}(b).

The strain-dependence of the resulting power factor PF is shown in \Ff{2}(c). Due to the weak impact of the thermopower, 
the behaviour of the power factor is dominated by the electrical conductivity dependence on the applied strain. 
At low temperatures and small charge carrier concentrations, such a behaviour has already been observed for biaxially strained 
silicon \cite{Hinsche:2011p15276}. However, the additional power-output described by the power factor is 
enhanced by 30\% under applied 1\% compressive strain for \PFpp and by 25\%  under applied 1\% tensile strain 
for \PFip. We note, that this low temperature and small doping case is not feasible for thermoelectric power 
generation, but could give inside in parasitic effects which play a role in metal-oxide-semiconductor devices.

Usual conditions for silicon-based thermoelectric applications, such as high temperature of \unit[900]{K} and 
large charge carrier concentrations $N = \unit[2\times 10^{21}]{cm^{-3}}$ are assumed in \Ffs{2}(d)-(f). 
At temperatures of 900 K the electronic band structure on a width $\pm \Delta E = \pm \unit[800]{meV}$ 
around the position of the chemical potential has to be considered, which makes a simplified description of
the electronic transport properties within a spherical band picture inaccurate. 
However, the dependence of the electrical conductivity (ref. \Ff{2}(d)) on the applied [111]-strain is almost 
preserved, even if accidentally. This is in contrast to biaxial strain in [001]-direction, were the 
strain-dependence induced by reoccupation of bands was suppressed under higher temperatures 
and dopings \cite{Hinsche:2011p15276,Dziekan:2007p1770}. 

In \Ff{2}(e) the anisotropic thermopower under trigonal distortion is shown. The cross-plane thermopower 
shows a monotonous increase from 85\% to 110\% of the unstrained thermopower value of $S_{_{\perp}} = \unit[115]{\mu V /K}$ varying from 
compressive to tensile strain. The in-plane component \Sip decreases to 85\% of the 
unstrained thermopower value under 1\% tensile strain. A very similar behaviour was 
found for the thermopower of biaxially strained silicon in [001]-direction \cite{Hinsche:2011p15276}. 
The compensation effects of the enhanced thermopower and decreased electrical conductivity and vice versa, 
are well known for thermoelectrics under strain \cite{Hinsche:2011p15276,Hinsche:2011p15707,Park:2010p11006}.
This scenario holds for [111]-strained silicon, too. 
In \Ff{2}(f) the shown anisotropic power factor in in-plane and cross-plane direction is always smaller 
than the power factor of the undistorted system, while \PFpp being at least stable under small values of 
compressive strain. We mention that the absolute values (cf right scales in \fs{2}(a) and (d)) of the electrical 
conductivity are increased remarkably compared to the low doping case as expected. Due to this, 
the absolute value of the power factor raises in its absolute value, but unfortunately does not show an 
enhancement due to mechanical strain in [111]-direction. 
Furthermore, compared to the low-doping/low-temperature regime (\Ff{2}(c)) the
power factor does not show noticeable anisotropy between in-plane
and cross-plane components.

\subsection{\label{hole} [111]-strained silicon: hole doping}

As it is well known, thermoelectric devices use two types of semiconductors, namely 
n-type and p-type, which are connected in series \cite{Sales:2002p6580}. Therefore, 
the influence of biaxial [111]-strain on hole-doped silicon is presented in \Ff{3} in the same 
way as done for the electron-doped case. In the low-doping/low-temperature 
regime an enormous enhancement for the cross-plane electrical conductivity \cpp under 
sufficient tensile strain can be found (cf \Ff{3}(a)), while the in-plane component \cip 
decreases more slightly under the same strain conditions. This behaviour can be 
linked to a changed subband structure. As mentioned before, strain 
lifts the degeneracy of LH and HH bands around the $\Gamma$ point and 
alters the curvature, that is the effective mass, of both bands. 
Under applied strain, the valence bands become highly 
anisotropic and a crossover between bands occurs 
so that they even loose their original LH and HH meaning\cite{Hinsche:2011p15276,Yu:2008p14181}. 
Extended discussions here on can be found in 
refs.~\cite{Sun:2007p14975,Thompson:2006p15708}. 

The thermopower of p-type silicon is shown in \Ff{3}(b). The thermopower in-plane 
and cross-plane decreases 
slightly under tensile, as well as under compressive strain. The anisotropy of the thermopower 
is moderate. As previously reported \cite{Hinsche:2011p15276,Pei:2011p15679}, the thermopower 
depends strongly on the number of occupied carrier pockets. A higher valley degeneracy at a 
fixed charge carrier concentration leads to an increased thermopower. As already mentioned the 
formerly degenerate HH and LH bands split under tensile and compressive strain. 
At low hole concentrations and low temperatures only the former HH (LH) band is occupied. 
This leads directly to a reduction of \Sip and \Spp. Consequently, the accompanied 
power factor (cf. \Ff{3}(c)) is also reduced in its maximal possible enhancement, but follows 
in principle the behaviour given by the electrical conductivity. 

For high temperatures and high hole concentrations the results are shown 
in \Ffs{3}(d)-(f). As the carrier concentration is raised by nearly six orders of magnitude with respect to the low 
doping case, the absolute 
value of the electrical conductivity raises in the same order (note the right scale of \Ff{3}(d)). 
Unfortunately, due to the higher band occupation and the broader smearing of the \textsc{Fermi-Dirac}-
distribution, effects of redistribution in strain-split bands do not play a role any more. 
Moreover, the effects of reduced and increased effective masses cancel each other leading to 
negligible change, in absolute values as well as in anisotropy, of the hole electrical conductivity.

For the hole thermopower shown in \Ff{3}(e) this behaviour is still valid. For the in-plane thermopower \Sip 
no significant influence of either compressive or tensile strained could be found. A minor dependence 
on applied [111]-strain is observed for \Spp. Here, the cross-plane thermopower decreases from 
$\unit[100]{\mu V/K}$ to $\unit[91]{\mu V/K}$ for strain values varying from 1\% compressive to 
1\% tensile strain.

Comprising the results for the electrical conductivity and thermopower, the resulting power factor 
under hole doping is shown in \Ff{3}(f). No evident influence of [111]-strain on the power factor 
could be found in the thermoelectrically relevant temperature and hole doping regime. Furthermore, the absolute 
value of the power factor is about 3-4 times smaller than in the comparable electron doped case, which 
is mainly caused by the higher hole scattering rate as assumed in Eq.~\ref{tau}.
\begin{figure*}[t]
\centering
\includegraphics[width=0.80\textwidth]{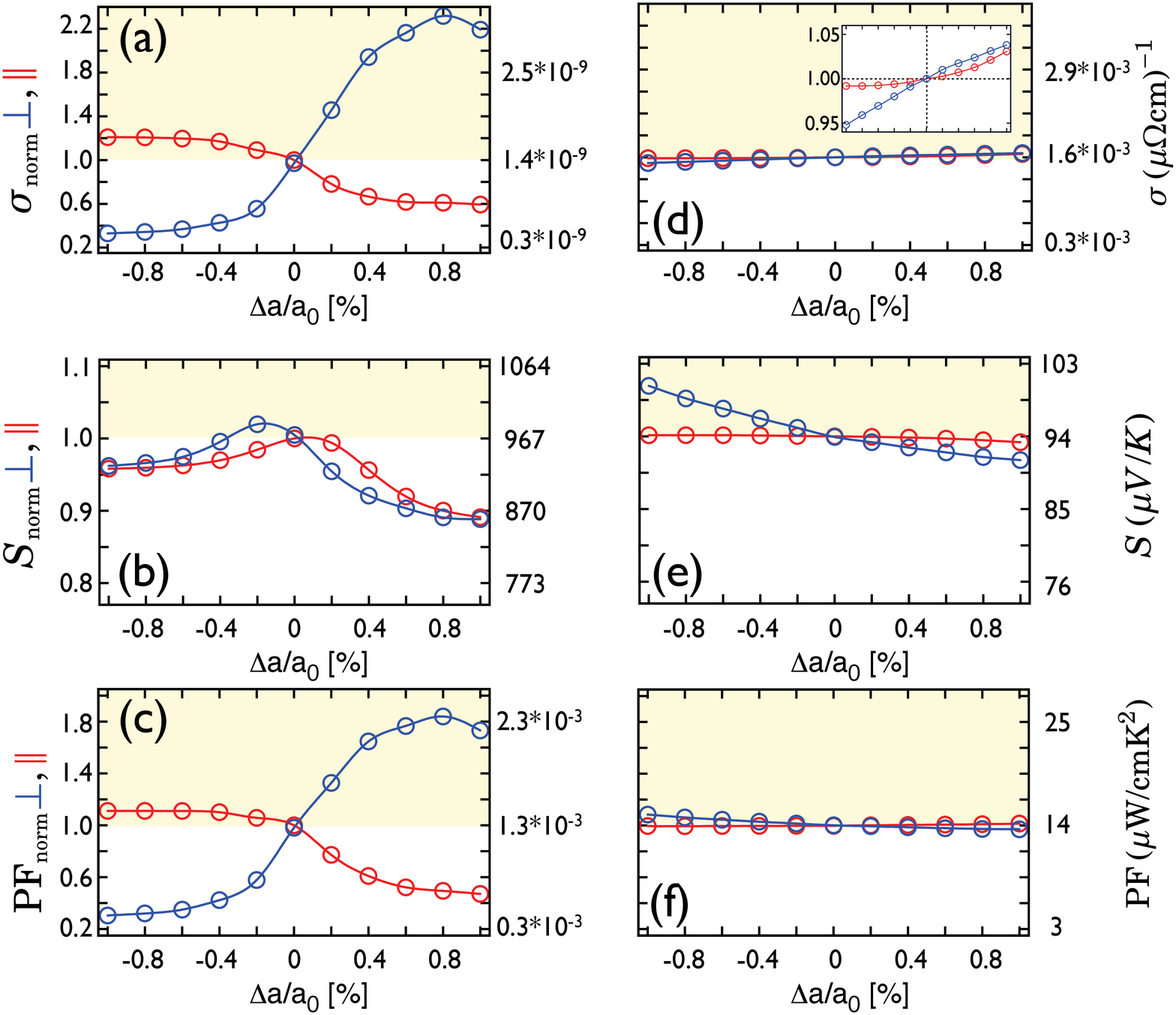}
\caption{\label{fig:3}(color online) Anisotropic thermoelectric transport properties of Si for fixed temperature 
and hole doping concentrations in dependence on compressive and tensile strain in [111]-direction. Left panels (a)-(c)) 
correspond to a hole doping of $\unit[2\times 10^{-8}]{e/atom}$ ($N = \unit[1\times 10^{15}]{cm^{-3}}$)
at a temperature of 100~K, while right panels (d)-(f) refer to a hole doping of $\unit[0.04]{e/atom}$ ($N = \unit[2\times 10^{21}]{cm^{-3}}$)
at a temperature of 900~K. On the left axis of each figure the relative value compared to the 
unstrained case is shown, while on the right axis the absolute values are given.}
\end{figure*}
At varying strain not only the absolute values of the thermoelectric properties change, 
but also the optimal charge carrier concentrations to obtain these maximized values. 
For biaxially [001]-strained silicon it was found, that the optimal doping 
range can change by a factor of two, while reducing the power factor up to 20\% 
if not adapting the charge carrier concentration \cite{Hinsche:2011p15276}. 
Therefore, the in-plane and cross-plane power factor under varying 
electron- and hole-doping, as well as varying [111]-strain is shown in \Ff{4}. The maximized 
power factor at optimal charge carrier concentration at a given strain state is emphasized by a black-dotted line in 
\Ffs{4}(a)-(d). Comparing silicon under electron doping (cf \Ffs{4}(a),(b)) 
and hole doping  (cf. \Ff{4}(c),(d)), it is obvious, that electron-doped silicon 
shows a much stronger variation of the optimal doping range. 
For the in-plane component \PFip and electron-doping the optimal carrier concentration 
decreases by about a factor 5, from $N = \unit[5\times 10^{21}]{cm^{-3}}$ to 
$N = \unit[1\times 10^{21}]{cm^{-3}}$ with the strain changing from 1\% compressive to 1\% tensile strain. 
For the corresponding cross-plane power factor \PFpp the optimal carrier concentration 
is about $N = \unit[1\times 10^{21}]{cm^{-3}}$ at 1\% compressive strain and 
increases by about a factor 5 under same strain conditions. At hole-doping no evident 
change of the optimal doping could be found at varying strain, while the absolute 
value depends only weakly on the applied strain (cf. \Ff{3}(f)). 
\begin{figure}
\centering
\includegraphics[width=0.58\textwidth]{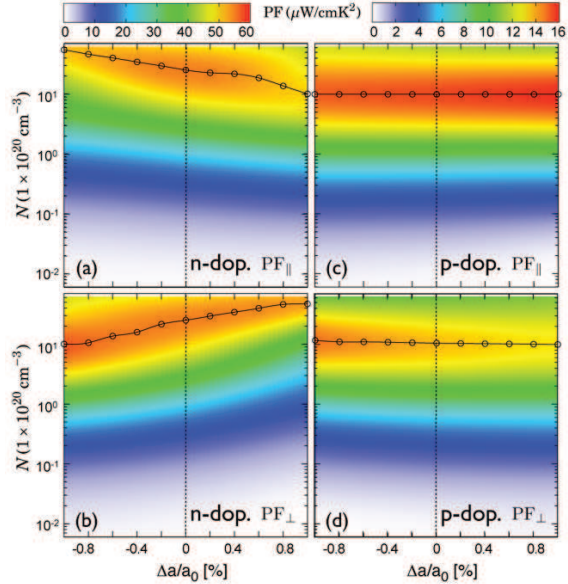}
\caption{(color online) In-plane and cross-plane power factor of Si at fixed temperature of 900K under electron (a,b) and 
hole doping (c,d) for varying charge carrier concentration and applied [111]-strain. The black circles emphasize 
the position of optimal doping at a certain strain state to maximize the power factor. 
Note the different scale for electron and hole doping.}
\label{fig:4}
\end{figure}

To summarize our findings, sections along the path of optimal electron doping are shown in 
\Ff{5}. With this, it is obvious that even with optimized doping no enhancement by tensile 
or compressive [111]-strain can be obtained for the in-plane power factor \PFip. For 
the cross-plane component \PFpp, which is more relevant for possible SL's, an increase 
of the power factor of about 4\% at 1\% compressive strain was found. 
To obtain this rather small enhancement the electron charge carrier concentration has to be reduced 
by about a factor of 2.5 compared to the unstrained case.
\begin{figure}
\centering
\includegraphics[width=0.48\textwidth]{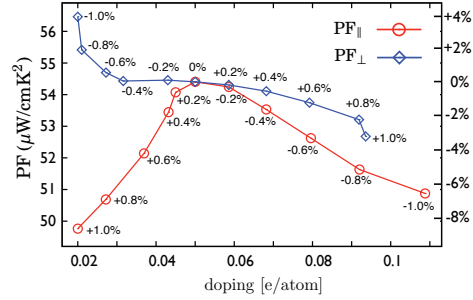}
\caption{(color online) Maximized in-plane and cross-plane power factor of [111]-strained Si at fixed temperature of 900K under optimal electron doping. 
The figure highlights the path of the black circles in fig.~\ref{fig:4}(a) and (b). $N = \unit[0.01]{e/atom}$ corresponds to 
$N = \unit[5\times 10^{20}]{cm^{-3}}$.}
\label{fig:5}  
\end{figure}
%

\subsection{\label{SiGe} Strained Si/Ge-SL on Si[111]}
By introducing the concepts of carrier pocket engineering \cite{Koga:1999p15445,Koga:1999p5363,Koga:2000p2542}
and phonon-glass/electron-crystal \cite{Slack_in_Rowe,Poudel:2008p7977} to semiconducting SLs 
an enormous leap forward to maximize the thermoelectric FOM was proposed. 
Indeed, several proofs-of-principle showed a remarkable enhancement of the FOM for 
thermoelectric semiconducting heterostructures 
\cite{Venkatasubramanian:2001p114,Harman:2002p5345,Koga:2000p2542,Hicks:1996p15693}.
With the thermal conductivity of SL's far below their alloy limit 
\cite{Venkatasubramanian:2000p7305,BorcaTasciuc:2000p15132,Chakraborty:2003p15514,Lee:1997p1545} and their constituents 
bulk values, a main task in optimizing the FOM is to enhance or at least to retain advantageous electronic properties of the 
bulk materials, so to say the power factor. For silicon-based SL's, carrier pocket engineering 
can be triggered by lattice strain. Using  Si$_{1-x}$/Ge$_{x}$ substrates in [111] and [100] orientation a 
$ZT=0.96$ and $0.24$, respectively, were predicted for strain-symmetrized\footnote{For an introduction to strain-symmetrized and 
strain-non-symmetrized Si/Ge superlattices we refer to the publications of \textsc{Kasper} \textit{et al.}\cite{Kasper:1988p15763,Kasper:1986p15761}.} 
Si(20\AA)Ge(20\AA) SL's at room temperature. 
The latter case was experimentally confirmed with $ZT=0.1$ at $N \approx \unit[1\times 10^{19}]{cm^{-3}}$ \cite{Koga:2000p2542}, 
which is nevertheless about a sevenfold enhancement relative to bulk Si \cite{Bux:2009p14985,Hinsche:2011p15276}. 
These experimental findings encourage further research for strain-non-symmetrized Si(20\AA)/Ge(20\AA) SLs in [001]-orientation 
and Si(15\AA)/Ge(40\AA) SLs in [111]-orientation, with $ZT=0.78$ and $1.25$, respectively, predicted at $T=\unit[300]{K}$ ~\cite{Koga:1999p15445}.

While being in principle possible for very thin films~\cite{Pearsall:1987p15762,Pearsall:1989p15764}, 
to our best knowledge state-of-the-art thin film technology 
does not enable strain-non-symmetrized SL's with satisfactory structural qualities and thicknesses
for thermoelectric applications so far \cite{Kasper:1988p15763,Kasper:1986p15761,Brunner:2002p15513,Kuan:1991p15765}. 
Nevertheless, in the following the thermoelectric transport properties of a strain-non-symmetrized 
Si(5\AA)/Ge(5\AA) SL in [111]-orientation will be discussed, as the largest enhancement of the FOM is expected here\cite{Koga:1999p15445}.

The used Si(5\AA)/Ge(5\AA) SL is represented by a hexagonal six-atom unit cell (ref. inset in \Ff{7}(a)) with point group symmetry $C_{3v}$ 
and a fixed in-plane lattice constant of $a=5.434${\AA} was used to simulate 
the bulk silicon substrate. Structural optimization of the atomic positions and the c-axis elongation of the unit cell 
were obtained using \textsc{VASP}~\cite{Kresse:1996p12346}. The distinct interlayer distances $\delta$ 
in [111]-direction were determined as $\delta_{Si_1-Si_2}=2.359${\AA}, $\delta_{Si_2-Si_3}=0.784${\AA}, $\delta_{Si_3-Ge_4}=2.409${\AA} , 
$\delta_{Ge_4-Ge_5}=0.929${\AA}, $\delta_{Ge_5-Ge_6}=2.470${\AA}, $\delta_{Ge_6-Si_1}=0.851${\AA} 
and are in good agreement with previous calculations \cite{Bass:1990p15515}. 
The ratio $\nicefrac{c}{a}=2.551$ shows an increase of the lattice constant in c-direction by about 4\% compared to bulk Si. 
This is clearly dictated by the fixed in-plane Si lattice constant and the larger volume of Ge. Due to this, the Ge layer 
can be seen as compressively strained in [111]-direction.

As has already been mentioned in literature~\cite{Bass:1990p15515,Salehpour:1990p15696}, the face centred cubic (fcc) 
L high symmetry point in the [111] direction folds onto the hexagonal A high symmetry point at a $\nicefrac{c}{a}$-ratio of $2.449$. 
In addition, the fcc X point is equivalent to the hexagonal M point, while the symmetry directions fcc $\Gamma$X 
and hexagonal $\Gamma$M are inequivalent. This is due the fcc X point lying in an adjoining Brillouin zone. 
Under biaxial [111]-strain two inequivalent sets of the eight L points occur. 
There are two points along c-axis in [111] direction of growth which fold onto the A 
point 
and those in the six directions equivalent to [$11\overline{1}$] hereafter noted as L.

\begin{figure}
\centering
\includegraphics[width=0.58\textwidth]{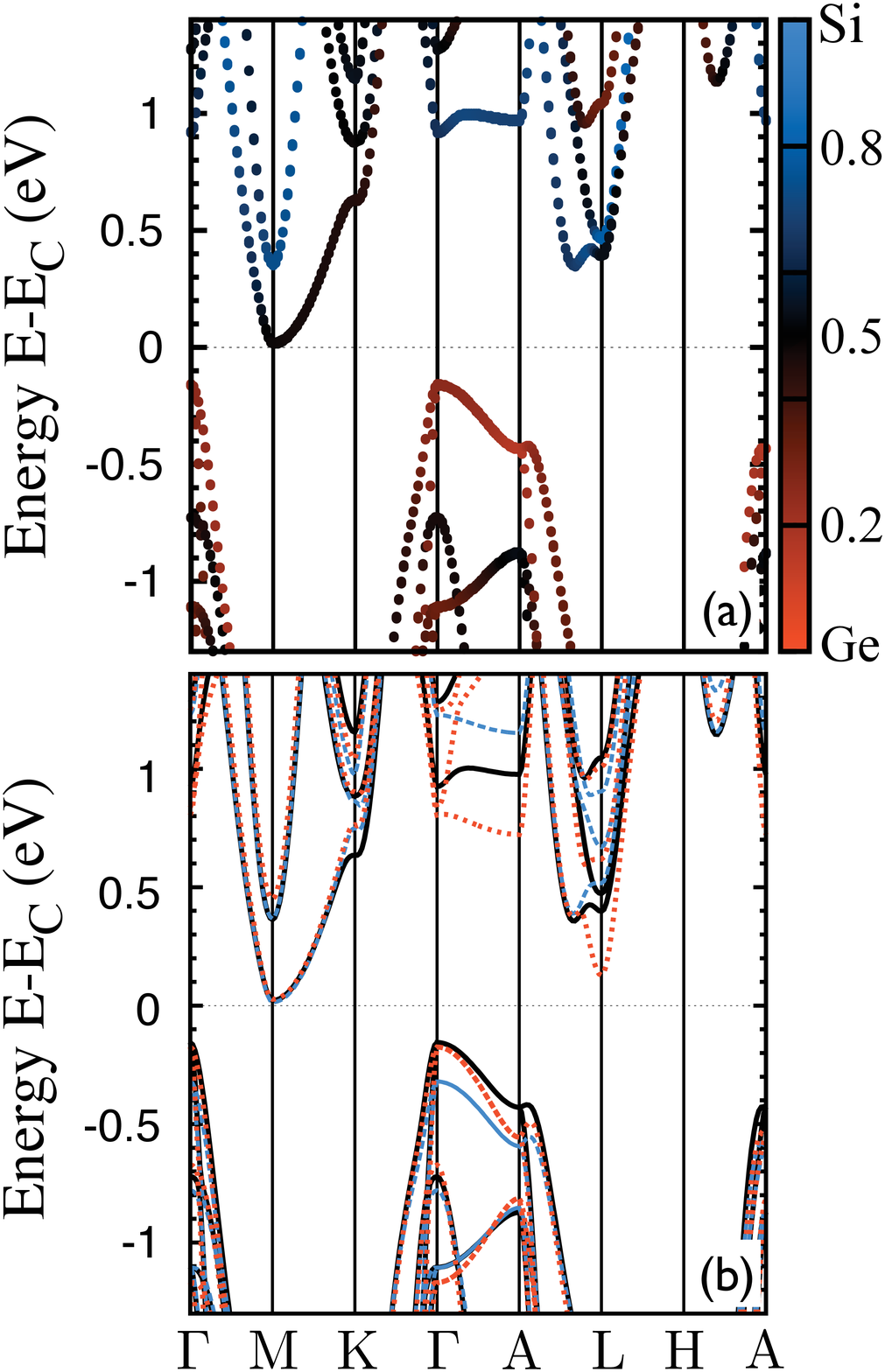}
\caption{(color online) (a) Band structure of the Si/Ge SL. The color code of the bands
 refers to the atomic character of the bands. Red dots refer to wavefunction character of pure germanium, while 
 blue dots refer to pure silicon like character. (b) band structure of silicon (blue, dashed lines), germanium 
 (red, dotted lines) and Si/Ge (black, solid line). All three configurations are calculated in 
 the lattice of Si/Ge. To allow comparison with (a), the bands are energetically matched at the CBM. 
 The figure has not been corrected for the band gap error.}
\label{fig:6}  
\end{figure}
In \Ff{6} the band structure for the fully relaxed Si/Ge-SL in the hexagonal unit-cell is shown. While in \Ff{6}(a) the 
site-resolved bands for the Si/Ge-SL are depicted, in \Ff{6}(b) the bands for pure Si on the fully-relaxed 
Si/Ge-positions are shown (blue, dashed lines), as well as all sites occupied by Ge (red, dotted lines) and 
the bands referring to the original Si/Ge-SL bands as shown in \Ff{6}(a) (black, solid lines).
Two main insights can be drawn. First, the VBM is located around the $\Gamma$ point and has an almost pure Ge character, 
given by the red dots in \Ff{6}(a). 
This is mainly due to the smaller band gap of Ge and the accompanied band offset between Ge and Si. Furthermore, 
the compressive biaxial strain in the Ge layers decreases the gap size and favours a direct band gap at $\Gamma$ instead 
of an indirect one between $\Gamma$ and L \cite{Tahini:2012p15819,Niquet:2009p15435,Zhang:2009p15487}.
Valence bands of mixed Si/Ge character come in to play around $\unit[0.65]{eV}$ below CBM and suggest the local 
indirect Si band gap between $\Gamma$ and M to be almost retained bulk like.
Second, the CBM is located at the M point and shows a strong mixing of Si and Ge character. Furthermore a 
strain-induced lifting of degeneracy occurs at the M point lowering a band of mixed character to the band edge and 
lifting a Si-like band upwards in energy. As only the Ge layers in the SL are compressively strained while the Si 
layers are nearly unstrained, the Ge L-point CBM valleys split into valleys located at A higher in energy 
and in L valleys lower in energy \cite{Koga:1999p15445,Zhang:2009p15487,Rideau:2006p15711}. These findings on 
the CBM and VBM characteristics are in agreement to experimental studies~\cite{Pearsall:1989p15764}.
With that the uncorrected band gap decreases to around $\unit[178]{meV}$, 
which is about 30\% of the uncorrected GGA gap for unstrained bulk Si. 
Furthermore, the effective masses at the CBM lower to M$_1$=$0.125 m_0$, 
M$_2$=$0.026 m_0$ and M$_3$=$0.010 m_0$ \footnote{The referring Eigenvectors were determined as 
$e_{1}=(0.8,0,0.6)$, $e_{2}=(0,1,0)$ and $e_{3}=(0.6,0,-0.8)$.}. 
Applying an effective mass approach \cite{Zahn:2011p15523} 
we find the conductivity anisotropy for energies near the CBM to \ratio{0.6}, which clearly prefers cross-plane transport 
under electron-doping.

The energy-dependent transport distribution function in the in-plane and cross-plane direction and their ratio are displayed in 
\Ff{7}(b) for the Si/Ge-SL and isotropic unstrained bulk Si. For the same systems the density of states 
are shown in \Ff{7}(a). 
As can be seen from \Ff{7}(b) unfortunately the conductivity anisotropy near 
the valence band edge strongly increases to values of 8 around $\unit[0.58]{eV}$ below CBM, clearly suppressing 
cross-plane electronic transport under p-type doping. This behaviour is largely due to the 
localization of the Ge like VBM states, in space as well as energy. As can be deduced from \Ff{6}(a) bands in the 
cross-plane direction (here $\Gamma$A) show pure Ge band character in an energy range of $\unit[0.18-0.74]{eV}$ 
below CBM. Clearly, in this energy range cross-plane conduction is suppressed by states localized in the Ge layers and almost 
vanishing in the Si layers leading to a strongly increased conductivity anisotropy. At an energy of $\unit[0.38]{eV}$ 
below CBM at $\Gamma$ a light band with strong Si/Ge mixed character appears (cf. black solid line in \Ff{6}(b)) leading 
to cross-plane transport through the Si and Ge layer forcing $\nicefrac{\sigma_{_{\|}}}{\sigma_{_{\perp}}}$ to 
decrease, while saturating for values above 1, still indicating a preferred in-plane transport under hole-doping.

\begin{figure}
\centering
\includegraphics[width=0.68\textwidth]{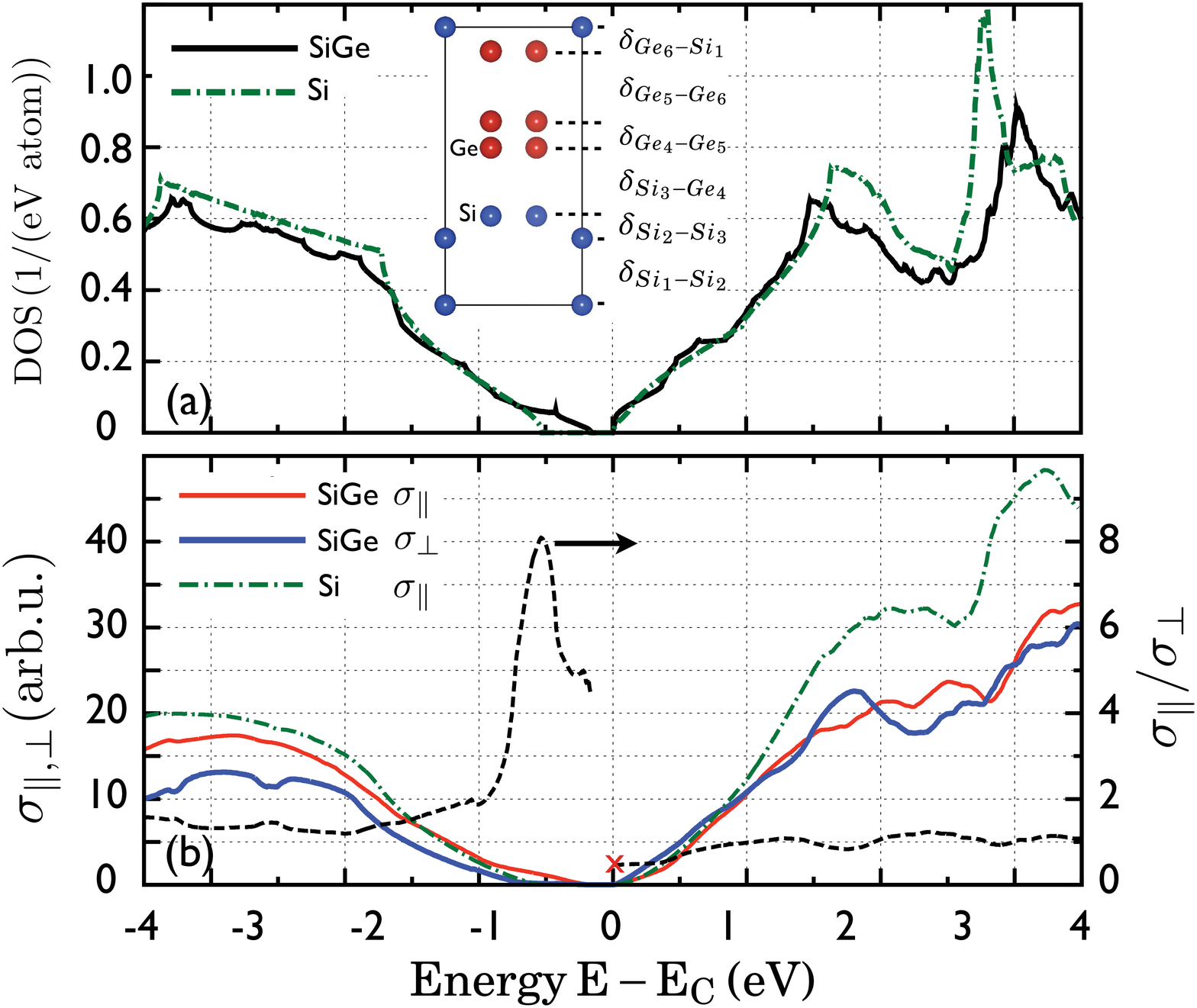}
\caption{(color online) (a) Density of states for bulk silicon (green, dashed-dotted line) and the Si/Ge SL (black, solid line). 
As an inset the hexagonal unit cell of the Si/Ge SL is shown~\cite{Momma}. Furthermore the interlayer distances are labelled, as referred to in the text.
(b) Electrical conductivity in dependence on the position of the chemical potential $\mu$ at zero temperature, shown for 
bulk silicon (green, dashed-dotted line) and the Si/Ge SL in the in-plane (red, solid line) and cross-plane direction (blue, solid line). 
The conductivity anisotropy (black, dashed line referring to the right axis) is stated for the Si/Ge SL. The cross at the CBM is 
the value obtained from an analytical effective mass approach.}
\label{fig:7} 
\end{figure}
In \Ff{8}(a),(b) the doping-dependent thermopower and power factor for the Si/Ge-SL are shown, respectively. As a 
comparison the reference values of bulk silicon are stated as black, dashed-dotted lines. 
We note, that a temperature dependency of the energy gap was introduced applying Eq.~\ref{teg}. 
For the Si/Ge-SL~\footnote{Here we adapted experimental data for Si$_{0.5}$Ge$_{0.5}$ alloys from Ref.~\cite{Braunstein:1958p15767}.} 
the parameters are chosen as 
$U_{\text{GGA}}=\unit[0.78]{eV}, \alpha=\unit[4.76 \times 10^{-4}]{\nicefrac{eV}{K}}$ and $\beta=\unit[395]{K}$. 
However, in the highly degenerate limit ($N > \unit[1\times 10^{20}]{cm^{-3}}$) the temperature dependence 
of the gap plays a negligible role even for temperatures above $\unit[900]{K}$. 

From \Ff{8}(a) it can be seen, that the thermopower in the Si/Ge-SL under electron doping (blue lines in the lower panel) 
is comparable to bulk silicon and follows a \textsc{Pisarenko} relation \cite{Ioffe:1960}. 
Under hole doping (red lines in the upper panel) the thermopower is suppressed compared to bulk silicon by 
about $\unit[80]{\mu V/K}$ for \Sipx and \Sppx in the relevant doping regime, which might be linked to a 
changed functional behaviour of the TDF $\mathcal{L}_{\perp, \|}^{(0,1)}(\mu, T)$. The latter can be deduced 
from \Ff{7}(b) where apparent differences in the functional behaviour of $\sigma_{_{\perp, \|}}$, which is proportional to 
$\mathcal{L}_{\perp, \|}^{(0)}(\mu, T)$, are visible especially in the valence bands of bulk Si and the Si/Ge-SL. 
The clear deviation of \Spp from the 
\textsc{Pisarenko} relation in the vicinity of $N \approx \unit[3\times 10^{21}]{cm^{-3}}$ is related to the strongly 
increased conductivity anisotropy $\unit[0.58]{eV}$ below VBM as shown in \Ff{7}(b). 
Here, the strong suppression of \cpp  causes a larger \Spp . 
Nevertheless, this slight enhancement of \Spp is not reflected in the power factor 
of the p-type Si/Ge-SL. 
As shown in \Ff{8}(b) \Sip, as well as \Spp are always 
smaller than the values for bulk Si under hole doping (red lines). Obviously, the suppressed electrical conductivity, especially in the 
cross-plane direction, is responsible for this result. A power factor of about $\unit[9]{\mu W/cm K^2}$ is 
found for \PFip at $N \approx \unit[3\times 10^{20}]{cm^{-3}}$, while in the cross-plane direction 
the same value can be stated at huge values of $N \approx \unit[3\times 10^{21}]{cm^{-3}}$, clearly 
evoked by the anomaly in the thermopower. More interesting is the case of electron doping (blue lines in \Ff{8}(a)). 
With the thermopower's behaviour almost bulk like and conductivity anisotropies $\nicefrac{\sigma_{_{\|}}}{\sigma_{_{\perp}}}$ 
below 1 an enhanced power factor in the required cross-plane direction is found. Compared to bulk silicon 
the \PFpp is enhanced by 10\% and reaches a value of $\unit[60]{\mu W/cm K^2}$ at an electron 
concentration of $N \approx \unit[7\times 10^{20}]{cm^{-3}}$. With that the optimal charge carrier concentration 
is four-times smaller compared to bulk Si. For the in-plane component \PFip almost no reduction can be seen, while 
the maximal value of $\unit[53]{\mu W/cm K^2}$ is shifted to slightly larger charge carrier concentrations. 
\begin{figure}
\centering
\includegraphics[width=0.48\textwidth]{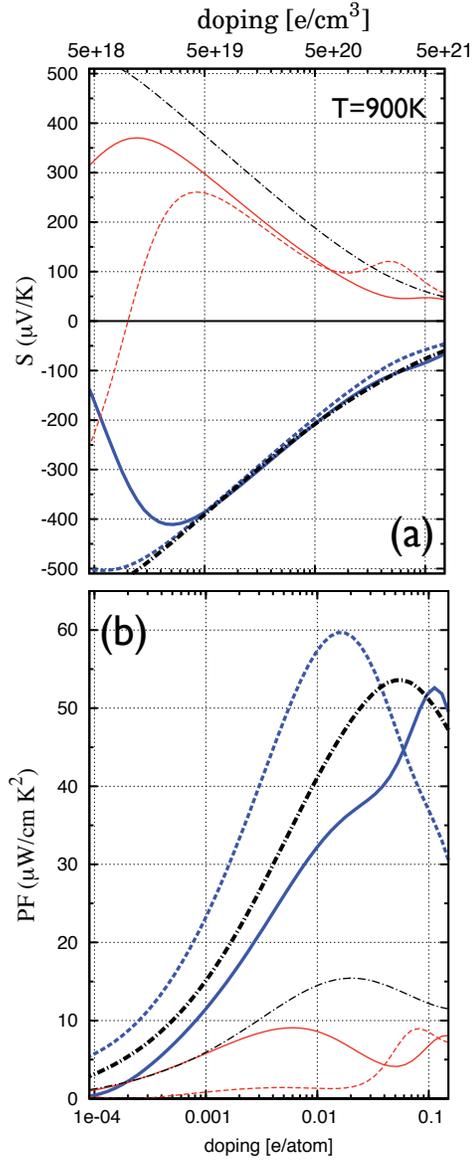}
\caption{(color online) (a) In-plane (solid lines) and cross-plane (dashed lines) doping-dependent thermopower 
at $\unit[900]{K}$ for the Si/Ge SL under electron (thick, blue lines) and hole-doping (thin, red lines). 
For comparison the values for bulk silicon are given (black, dashed-dotted lines). 
(b) In-plane (solid lines) and cross-plane (dashed lines) doping-dependent power factor 
at $\unit[900]{K}$ for the Si/Ge SL under electron (thick, blue lines) and hole-doping (thin, red lines). 
Again, for comparison the values for bulk silicon are given (black, dashed-dotted lines).}
\label{fig:8}  
\end{figure}
\subsection{\label{FOM} Towards the figure of merit}
\begin{figure}
\centering
\includegraphics[width=0.68\textwidth]{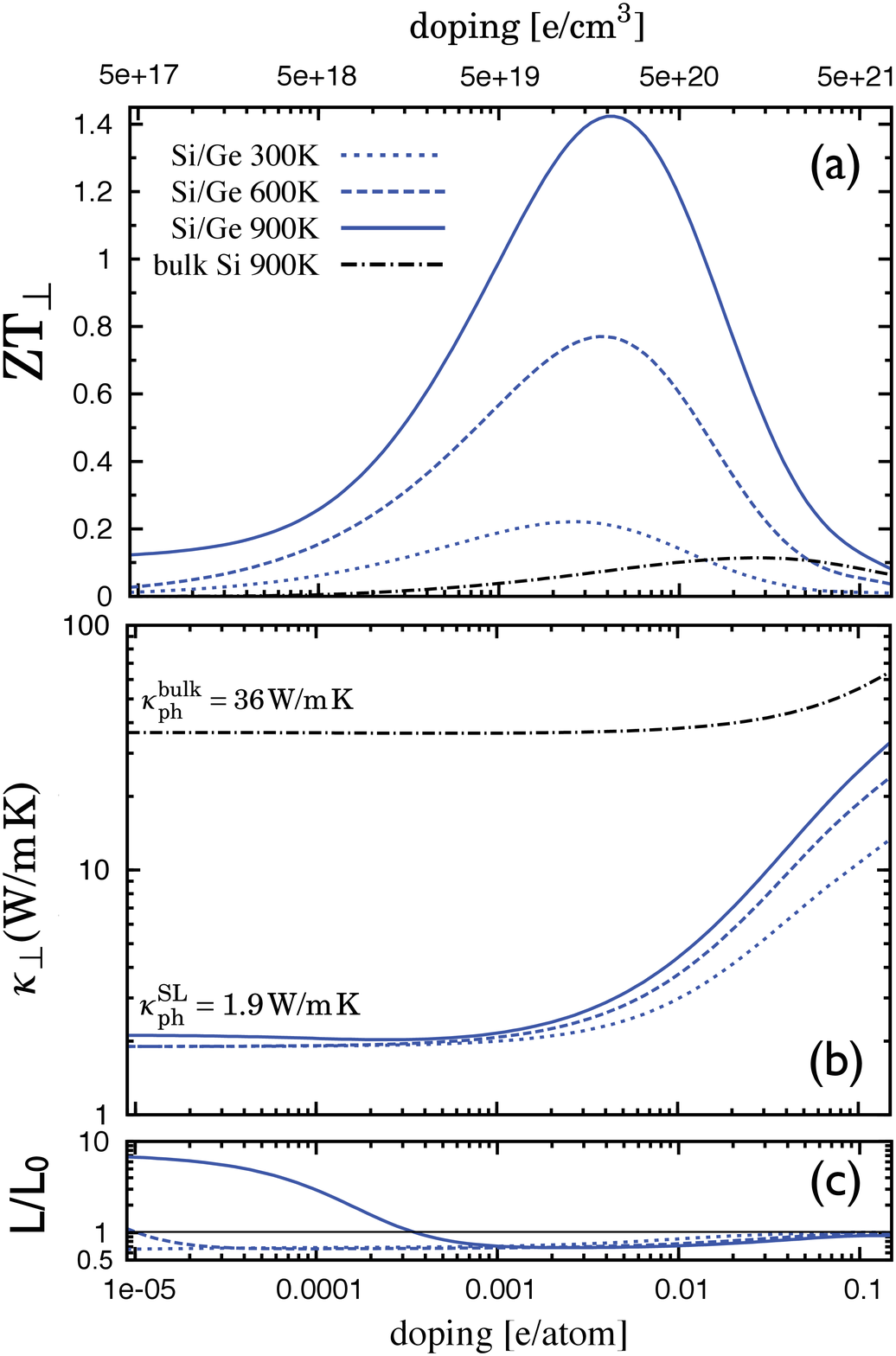}
\caption{(color online) (a) Doping-dependent cross-plane figure of merit ZT$_{\perp}$ of the Si/Ge SL (blue lines)
and bulk Si (black lines) under electron-doping at different temperatures. (b) the total thermal conductivity $\kappa_{\perp}$ 
in cross-plane direction for the Si/Ge SL (blue lines) and bulk Si (black lines) under electron-doping at different temperatures. 
While the electronic part $\kappa_{el}{_{\perp}}$ was calculated, the lattice part $\kappa_{ph}{_{\perp}}$ was estimated from experiments 
\cite{Bux:2009p14985,Lee:1997p1545} and is constant for varying charge carrier concentration. 
(c) the calculated cross-plane Lorenz function $L_{\perp}=\kappa_{el}{_{\perp}} \cdot \left( \sigma_{\perp} T \right)^{-1}$
related to the metallic limit $L_0=\unit[2.44\times 10^{-8}]{W \Omega /K^2}$.}
\label{fig:9}  
\end{figure}
In \Ff{9}(a) the FOM in cross-plane direction for the Si/Ge-SL (blue lines) and the more promising electron-doped case is shown. 
Different temperatures are chosen to demonstrate the evaluation of maximal ZT and the range of optimal charge carrier concentration. 
As a comparison the FOM for bulk Si is shown as black dashed-dotted line, too. 
To present results for the FOM, knowledge of the thermal conductivity is relevant. For this purpose, the electronic part of 
the thermal conductivity $\kappa_{el}{_{\perp}}$ was calculated applying Eq.~\ref{kel}, while the lattice part $\kappa_{ph}{_{\perp}}$ was 
taken from experiment. Here, $\kappa_{ph}{_{\perp}}=\unit[36]{W/m K}$ \cite{Bux:2009p14985} was used for bulk Si at $\unit[900]{K}$, while 
$\kappa_{ph}{_{\perp}}=\unit[1.9]{W/m K}$ was used for the Si/Ge-SL \cite{Lee:1997p1545}. With the latter being rather optimistic, 
since smaller than the expected nano-alloy limit of $\unit[2.5]{W/m K}$ \cite{Wang:2008p15709}, but achievable in Si/Ge-SL \cite{BorcaTasciuc:2000p15132,Lee:1997p1545,Pernot:2010p14944}. 
However, from \Ff{9}(b) one can deduce that for thermoelectric reliable charge carrier concentrations above 
$\unit[4\times 10^{20}]{cm^{-3}}$ the electronic contribution to the total thermal conductivity dominates over its lattice part. 
At $\unit[900]{K}$ for $N \approx \unit[1.5\times 10^{20}]{cm^{-3}}$ 
the electronic part amounts to $\unit[0.73]{W/m K}$ clearly smaller than the lattice part. This 
contribution increases significantly at higher temperatures and charge carrier concentrations omitting higher 
absolute values of the FOM. However, ZT$_{\perp}$ above unity can be reached for operating temperatures higher 
than $\unit[750]{K}$ and electron-doping $N \approx \unit[1.5-3\times 10^{20}]{cm^{-3}}$. 

At room-temperature a ZT$_{\perp}\approx 0.2$ is achieved in a broad doping range of $N \approx \unit[5-25\times 10^{19}]{cm^{-3}}$. 
Besides this value being an order of magnitude higher than the bulk Si value of ZT$ \approx 0.01$ 
\cite{Bux:2009p14985,Koga:1999p15445,Hinsche:2011p15276}, 
it is still less than the postulated values of ZT$_{\perp}=0.96$ and ZT$_{\perp}=1.25$ 
by Koga \textit{et al.}\cite{Koga:1999p15445} for strain-symmetrized 
and strain-non-symmetrized Si/Ge-SL, respectively. With that, and using a rather conservative 
value of $\kappa_{ph}{_{\perp}}=\unit[7.3]{W/m K}$ for their estimations, they expect 
enormous power factors of PF$_{\perp} \approx \unit[250]{\mu W/cm K^2}$ and PF$_{\perp} \approx \unit[340]{\mu W/cm K^2}$ 
for the strain-optimized Si/Ge-SL in [111]-direction. 
We found PF$_{\perp} \approx \unit[15]{\mu W/cm K^2}$ at $\unit[300]{K}$ and $N \approx \unit[1\times 10^{20}]{cm^{-3}}$ 
for the Si/Ge-SL. 
Even though a convergence of carrier pockets is not fully achieved in our superlattice, power factors 
beyond $\unit[200]{\mu W/cm K^2}$ seem to be very sophisticated, as state-of-the art power factors 
near or above room-temperature are well below $\unit[100]{\mu W/cm K^2}$ 
\cite{Venkatasubramanian:2001p114,Jalan:2010p14038,Okinaka:2006p15720}. A benefit from thermoionic 
emission even at moderate temperatures can not be expected in Si/Ge-SL \cite{Vashaee:2007p15721,Shakouri:1997p15266,Vashaee:2004p6592}. 
We note, that at very low temperatures below $\unit[10]{K}$ huge PF of about $\unit[100-1000]{\mu W/cm K^2}$ 
were reported for bulk Fe$_{2}$Sb$_{2}$ and ZnO$_{1-x}$Se$_{x}$ ~\cite{Sun:2009p15718,Lee:2010p6879}.

As an addition in \Ff{9}(c) 
the doping-dependent Lorenz function $L_{\perp}=\kappa_{el}{_{\perp}} \cdot \left( \sigma_{\perp} T \right)^{-1}$ 
as defined via equations (\ref{Seeb}) and (\ref{kel}) is presented. From \Ff{9}(c) it is obvious that the Lorenz number 
$L_{\perp}$ can be substantially different from the metallic limit $L_0$. Nevertheless, for very large charge carrier concentrations 
and the chemical potential located deep inside the conduction band, $L_{\perp}$ almost coincides with $L_0$. 
At intermediate and thermoelectric relevant charge carrier concentrations of $N \approx \unit[5-50\times 10^{19}]{cm^{-3}}$ 
$L_{\perp}$ can be much smaller than $L_0$. For $\unit[900]{K}$ and $N \approx \unit[1.25\times 10^{20}]{cm^{-3}}$ 
we find a minimal value of $L_{\perp} \approx 0.685 L_0$. At smaller charge carrier concentrations $L_{\perp}$ rapidly 
increases and reaches $L_{\perp} \approx 7 L_0$ for electron charge carrier concentrations of 
$N < \unit[1\times 10^{18}]{cm^{-3}}$ in the intrinsic doping regime. At decreasing temperatures minimal values 
of the Lorenz function are obtained at much smaller charge carrier concentrations. Furthermore, the maximal values 
of $L$ shifts to smaller charge carrier concentrations, too, and can reach huge values of $L$ at very low temperatures 
and charge carrier concentrations. The effect, which is responsible for the suppression of the Lorenz function to values below the
metallic limit $L_0$ is termed \textit{bipolar
thermodiffusion effect}~\cite{Tritt:2004p15755,Chaput:2005p1405,FlageLarsen:2011p15689} and maximizes 
for positions of the chemical potential near the band edges. 
However, a Lorenz function $L \neq L_0$ can
have consequences for the determination of the thermal
conductivity. The Lorenz factor is generally used to separate $\kappa_{el}$ and $\kappa_{ph}$. 
At thermoelectric advisable charge
carrier concentrations applying the metallic value $L_0$ to
determine the lattice thermal conductivity could lead to
an overestimation of the electronic thermal conductivity,
and consequently to an underestimation of the lattice
contribution.

\section{Conclusions and Outlook}
With the presented results we have shown, that strain in [111]-direction is not sufficient to 
significantly enhance the thermoelectric transport properties in bulk Si for energy harvesting applications. 
In the low-temperature and low-doping case large enhancements were found at tensile 
strain for \PFip (electron-doping) and \PFpp (hole-doping) and under compressive strain for 
\PFpp (electron-doping). This could have a negative impact for metal-oxide-semiconductor 
devices involving [111]-strained Si. Here, in the low-temperature and low-doping regime, 
small temperature gradients in the devices could lead to an additional parasitic electrical power, 
which can be way larger than expected from unstrained bulk. 

Enhancements found in the high-temperature and high doping 
regime were distinct smaller. Here slight enhancements of 5\% for \PFpp were found under compressive 
strain. The more interesting is, that the power factor is robust against [111]-strain, especially 
under hole-doping. Thus, thermoelectric SLs based on [111]-strained Si could provide an enhanced FOM, 
as $\kappa_{ph}$ is most likely reduced in SLs. 
We note, that due the to high bulk thermal conductivity and 
the only modest gain in the power factor by 
[111]-strain engineering, bulk silicon remains an unfavourable thermoelectric, even if 
the electronic transport properties are strain optimized. 
However, from comparison with earlier studies on biaxially [001]-strained 
silicon~\cite{Hinsche:2011p15276}, we confirm that 
strain in [111]-direction, e.g. in silicon-based SLs, should be preferred, as the carrier pocket 
degeneracy is retained and therefore the thermopower and power factor can be maximized.

To deal with that, we investigated the anisotropic thermoelectric 
transport of an [111]-oriented Si/Ge superlattice. 
At a first glance we have shown that no degradation of the electronic transport 
by the heterostructure is expected for electron doping, while even showing an enhancement 
of 10\% in \PFpp compared to bulk Si. Assuming a decrease in lattice thermal conductivity 
a large enhancement in ZT to 0.2 and 1.4 is achieved at $\unit[300]{K}$ and $\unit[900]{K}$, respectively. 
Under hole doping the electronic transport in the Si/Ge-SL is heavily suppressed due to 
quantum well effects. Here the cross-plane power factor \PFpp is expected to show only 
around 50\% of the bulk maximal value, leading to small ZT values.

\begin{ack} 
This work was supported by the Deutsche 
Forschungsgemeinschaft, SPP 1386 `Nanostrukturierte Thermoelektrika: 
Theorie, Modellsysteme und kontrollierte Synthese'. N. F. Hinsche is 
member of the International Max Planck Research School for Science 
and Technology of Nanostructures.
\end{ack}

\section*{References}
\bibliographystyle{unsrt}
\bibliography{paper.bib}

\begin{thebibliography}{10}

\bibitem{Seebeck}
T.~J. Seebeck.
\newblock {\"Uber die magnetische Polarisation der Metalle und Erze durch
  Temperatur-Differenz}.
\newblock {\em Annalen der Physik}, 82:253--286, 1826.

\bibitem{Ioffe1958}
A.~F. Ioffe.
\newblock The revival of thermoelectricity.
\newblock {\em Scientific American}, 199:31--37, 1958.

\bibitem{Majumdar:2004p6568}
A~Majumdar.
\newblock {Thermoelectricity in Semiconductor Nanostructures}.
\newblock {\em Science}, 303(5659):777, 2004.

\bibitem{Bottner:2006p2812}
H~B{\"o}ttner, G~Chen, and R~Venkatasubramanian.
\newblock {Aspects of Thin-Film Superlattice Thermoelectric Materials, Devices,
  and Applications}.
\newblock {\em MRS bulletin}, 31:211, 2006.

\bibitem{Tritt:2006p15694}
T.M Tritt and MA~Subramanian.
\newblock Thermoelectric materials, phenomena, and applications: A bird's eye
  view.
\newblock {\em MRS bulletin}, 31(03):188--198, 2006.

\bibitem{Sales:2002p6580}
BC~Sales.
\newblock Thermoelectric materials: Smaller is cooler.
\newblock {\em Science}, 295(5558):1248, 2002.

\bibitem{Venkatasubramanian:2001p114}
R~Venkatasubramanian, E~Siivola, and T~Colpitts.
\newblock Thin-film thermoelectric devices with high room-temperature figures
  of merit.
\newblock {\em Nature}, 413:597, 2001.

\bibitem{Harman:2002p5345}
T~Harman, P~Taylor, M~Walsh, and B~LaForge.
\newblock Quantum dot superlattice thermoelectric materials and devices.
\newblock {\em Science}, 297:2229, 2002.

\bibitem{Dresselhaus:2007p2775}
M~Dresselhaus, G~Chen, M~Tang, and R~Yang.
\newblock New directions for low-dimensional thermoelectric materials.
\newblock {\em Advanced Materials}, 19:1, 2007.

\bibitem{Vining:2008p9416}
CB~Vining.
\newblock Materials science: Desperately seeking silicon.
\newblock {\em Nature}, 451(7175):132--133, 2008.

\bibitem{Hochbaum:2008p6569}
A~Hochbaum, R~Chen, R~Delgado, and W~Liang.
\newblock Enhanced thermoelectric performance of rough silicon nanowires.
\newblock {\em Nature}, 451:163, 2008.

\bibitem{Boukai:2008p14967}
AI~Boukai, Y~Bunimovich, J~Tahir-Kheli, JK~Yu, WA~Goddard Iii, and JR~Heath.
\newblock Silicon nanowires as efficient thermoelectric materials.
\newblock {\em Nature}, 451(7175):168--171, 2008.

\bibitem{Bux:2009p14985}
Sabah~K Bux, Richard~G Blair, Pawan~K Gogna, Hohyun Lee, Gang Chen, Mildred~S
  Dresselhaus, Richard~B Kaner, and Jean-Pierre Fleurial.
\newblock {Nanostructured Bulk Silicon as an Effective Thermoelectric
  Material}.
\newblock {\em Adv. Funct. Mater.}, 19(15):2445--2452, 2009.

\bibitem{Tang:2010p15127}
J~Tang, H.T Wang, D.H Lee, M~Fardy, Z~Huo, T.P Russell, and P~Yang.
\newblock {Holey Silicon as an Efficient Thermoelectric Material}.
\newblock {\em Nano Lett}, 10:4279, 2010.

\bibitem{Koga:1999p15445}
T~Koga, X~Sun, SB~Cronin, and MS~Dresselhaus.
\newblock Carrier pocket engineering applied to strained si/ge superlattices to
  design useful thermoelectric materials.
\newblock {\em Applied Physics Letters}, 75:2438, 1999.

\bibitem{Koga:1999p5363}
T~Koga, T~Harman, S~Cronin, and M~Dresselhaus.
\newblock {Mechanism of the enhanced thermoelectric power in (111)-oriented
  n-type PbTe/Pb$_{1-x}$Eu$_x$Te multiple quantum wells}.
\newblock {\em Phys. Rev. B}, 60(20):14286, 1999.

\bibitem{Koga:2000p2542}
T~Koga, S~Cronin, M~Dresselhaus, and J~Liu.
\newblock {Experimental proof-of-principle investigation of enhanced ZT in
  (001) oriented Si/Ge superlattices}.
\newblock {\em Applied Physics Letters}, 77(10), 2000.

\bibitem{Hinsche:2011p15276}
N~Hinsche, I~Mertig, and P~Zahn.
\newblock Effect of strain on the thermoelectric properties of silicon: an ab
  initio study.
\newblock {\em J. Phys.: Condens. Matter}, 23:295502, 2011.

\bibitem{Giannozzi:2009p14969}
Paolo Giannozzi, Stefano Baroni, Nicola Bonini, Matteo Calandra, Roberto Car,
  Carlo Cavazzoni, Davide Ceresoli, Guido Chiarotti, Matteo Cococcioni, Ismaila
  Dabo, Andrea~Dal Corso, Stefano~De Gironcoli, Stefano Fabris, Guido Fratesi,
  Ralph Gebauer, Uwe Gerstmann, Christos Gougoussis, Anton Kokalj, Michele
  Lazzeri, Layla Martin-Samos, Nicola Marzari, Francesco Mauri, Riccardo
  Mazzarello, Stefano Paolini, Alfredo Pasquarello, Lorenzo Paulatto, Carlo
  Sbraccia, Sandro Scandolo, Gabriele Sclauzero, Ari Seitsonen, Alexander
  Smogunov, Paolo Umari, and Renata Wentzcovitch.
\newblock Quantum espresso: a modular and open-source software project for
  quantum simulations of materials.
\newblock {\em J. Phys.: Condens. Matter}, 21:395502, 2009.

\bibitem{Mertig:1999p12776}
I~Mertig.
\newblock Transport properties of dilute alloys.
\newblock {\em Reports on Progress in Physics}, 62:237--276, 1999.

\bibitem{Hinsche:2011p15707}
N~Hinsche, B~Yavorsky, I~Mertig, and P~Zahn.
\newblock {Influence of strain on anisotropic thermoelectric transport in
  Bi$_{2}$Te$_{3}$ and Sb$_{2}$Te$_{3}$}.
\newblock {\em Physical Review B}, 84(16):165214, 2011.

\bibitem{Zahn:2011p15523}
Peter Zahn, Nicki Hinsche, B~Yavorsky, and Ingrid Mertig.
\newblock {Bi$_2$Te$_3$: implications of the rhombohedral k-space texture on
  the evaluation of the in-plane/out-of-plane conductivity anisotropy}.
\newblock {\em J. Phys.: Condens. Matter}, 23:505504, 2011.

\bibitem{Perdew:1996p14792}
JP~Perdew, K~Burke, and M~Ernzerhof.
\newblock Generalized gradient approximation made simple.
\newblock {\em Phys. Rev. Lett.}, 77(18):3865--3868, 1996.

\bibitem{Corso:2005p8612}
A~Corso and A~Conte.
\newblock Spin-orbit coupling with ultrasoft pseudopotentials: Application to
  au and pt.
\newblock {\em Phys. Rev. B}, 71:115106, 2005.

\bibitem{Yu:2008p14181}
D~Yu, Y~Zhang, and F~Liu.
\newblock {First-principles study of electronic properties of biaxially
  strained silicon: Effects on charge carrier mobility}.
\newblock {\em Physical Review B}, 78(24):245204, 2008.

\bibitem{Bouhassoune:2009p6886}
M~Bouhassoune and A~Schindlmayr.
\newblock Electronic structure and effective masses in strained silicon.
\newblock {\em physica status solidi (c)}, 7(2):460--463, 2009.

\bibitem{Kresse:1996p12346}
G~Kresse and J~Furthm{\"u}ller.
\newblock Efficient iterative schemes for ab initio total-energy calculations
  using a plane-wave basis set.
\newblock {\em Physical Review B}, 54(16):11169, 1996.

\bibitem{Godby:1988p14795}
RW~Godby, M~Schl{\"u}ter, and LJ~Sham.
\newblock Self-energy operators and exchange-correlation potentials in
  semiconductors.
\newblock {\em Physical Review B}, 37(17):10159--10175, 1988.

\bibitem{Varshni:1967p14976}
YP~Varshni.
\newblock Temperature dependence of the energy gap in semiconductors.
\newblock {\em Physica}, 34(1):149--154, 1967.

\bibitem{Vojta:1992p1395}
T~Vojta, I~Mertig, and R~Zeller.
\newblock Calculation of the residual resistivity and the thermoelectric power
  of sp impurities in silver.
\newblock {\em Phys. Rev. B}, 46(24):16761, 1992.

\bibitem{Thonhauser:2004p14960}
T~Thonhauser, TJ~Scheidemantel, and JO~Sofo.
\newblock Improved thermoelectric devices using bismuth alloys.
\newblock {\em Applied Physics Letters}, 85:588, 2004.

\bibitem{Singh:2010p14285}
David~J Singh.
\newblock {Doping-dependent thermopower of PbTe from Boltzmann transport
  calculations}.
\newblock {\em Physical Review B}, 81(19):195217, 2010.

\bibitem{Scheidemantel:2003p14961}
TJ~Scheidemantel, C~Ambrosch-Draxl, T~Thonhauser, JV~Badding, and JO~Sofo.
\newblock Transport coefficients from first-principles calculations.
\newblock {\em Physical Review B}, 68(12):125210, 2003.

\bibitem{Jacoboni:1977p14945}
C~Jacoboni, C~Canali, G~Ottaviani, and A~Alberigi Quaranta.
\newblock A review of some charge transport properties of silicon.
\newblock {\em Solid-State Electronics}, 20(2):77--89, 1977.

\bibitem{Dziekan:2007p1770}
T~Dziekan, P~Zahn, V~Meded, and S~Mirbt.
\newblock Theoretical calculations of mobility enhancement in strained silicon.
\newblock {\em Physical Review B}, 75(19):195213, 2007.

\bibitem{Roldan:2012p15129}
J~B Rold{\'a}n, F~G{\'a}miz, J~A L{\'o}pez‐Villanueva, and J~E Carceller.
\newblock {A Monte Carlo study on the electron‐transport properties of
  high‐performance strained‐Si on relaxed Si$_1−x$Ge$x$ channel MOSFETs}.
\newblock {\em Journal of Applied Physics}, 80(9):5121, 1996.

\bibitem{Mahan:1996p508}
G~Mahan and JO~Sofo.
\newblock The best thermoelectric.
\newblock {\em Proceedings of the National Academy of Sciences}, 93:7436, 1996.

\bibitem{Lehmann:1972p14972}
G~Lehmann and M~Taut.
\newblock {On the Numerical Calculation of the Density of States and Related
  Properties}.
\newblock {\em physica status solidi (b)}, 54(2):469--477, 1972.

\bibitem{Zahn:1995p14971}
P~Zahn, I~Mertig, M~Richter, and H~Eschrig.
\newblock {Ab Initio Calculations of the Giant Magnetoresistance}.
\newblock {\em Phys. Rev. Lett.}, 75(16):2996--2999, 1995.

\bibitem{Mertig:1987p5922}
Ingrid Mertig, Eberhard Mrosan, and Paul Ziesche.
\newblock {\em Multiple scattering theory of point defects in metals:
  Electronic properties}.
\newblock B.G. Teubner, Leipzig, 1987.

\bibitem{Momma}
Koichi Momma and Fujio Izumi.
\newblock {{\it VESTA3} for three-dimensional visualization of crystal,
  volumetric and morphology data}.
\newblock {\em Journal of Applied Crystallography}, 44(6):1272--1276, Dec 2011.

\bibitem{Baykan:2012p14974}
Mehmet~O Baykan, Scott~E Thompson, and Toshikazu Nishida.
\newblock Strain effects on three-dimensional, two-dimensional, and
  one-dimensional silicon logic devices: Predicting the future of strained
  silicon.
\newblock {\em Journal of Applied Physics}, 108(9):093716, 2010.

\bibitem{Sun:2007p14975}
Guangyu Sun, Yongke Sun, Toshikazu Nishida, and Scott~E Thompson.
\newblock Hole mobility in silicon inversion layers: Stress and surface
  orientation.
\newblock {\em Journal of Applied Physics}, 102(8):084501, 2007.

\bibitem{Boykin:2007p15488}
Timothy Boykin, Neerav Kharche, and Gerhard Klimeck.
\newblock {Brillouin-zone unfolding of perfect supercells having nonequivalent
  primitive cells illustrated with a Si∕Ge tight-binding parameterization}.
\newblock {\em Phys. Rev. B}, 76(3):035310, 2007.

\bibitem{Niquet:2009p15435}
Y~Niquet, D~Rideau, C~Tavernier, H~Jaouen, and X~Blase.
\newblock Onsite matrix elements of the tight-binding hamiltonian of a strained
  crystal: Application to silicon, germanium, and their alloys.
\newblock {\em Physical Review B}, 79(24):245201, 2009.

\bibitem{Park:2010p11006}
Min~Sik Park, Jung-Hwan Song, Julia~E Medvedeva, Miyoung Kim, In~Gee Kim, and
  Arthur~J Freeman.
\newblock Electronic structure and volume effect on thermoelectric transport in
  p -type bi and sb tellurides.
\newblock {\em Phys. Rev. B}, 81(15):155211, 2010.

\bibitem{Thompson:2006p15708}
S.E Thompson, G~Sun, Y.S Choi, and T~Nishida.
\newblock {Uniaxial-process-induced strained-Si: extending the CMOS roadmap}.
\newblock {\em IEEE Transactions on Electron Devices}, 53(5):1010--1020, 2006.

\bibitem{Pei:2011p15679}
Yanzhong Pei, Xiaoya Shi, Aaron LaLonde, Heng Wang, Lidong Chen, and G.~Jeffrey
  Snyder.
\newblock Convergence of electronic bands for high performance bulk
  thermoelectrics.
\newblock {\em Nature}, 473(7345):66--69, 2011.

\bibitem{Slack_in_Rowe}
Glen~A. Slack.
\newblock {\em {CRC Handbook of Thermoelectrics}}, chapter~34, page 407.
\newblock CRC Press, Boca Raton, 1995.

\bibitem{Poudel:2008p7977}
B~Poudel, Q~Hao, Y~Ma, Y~Lan, A~Minnich, B~Yu, X~Yan, D~Wang, A~Muto, and
  D~Vashaee.
\newblock High-thermoelectric performance of nanostructured bismuth antimony
  telluride bulk alloys.
\newblock {\em Science}, 320(5876):634, 2008.

\bibitem{Hicks:1996p15693}
L~Hicks, T~Harman, X~Sun, and M~Dresselhaus.
\newblock Experimental study of the effect of quantum-well structures on the
  thermoelectric figure of merit.
\newblock {\em Physical Review B}, 53(16):R10493--R10496, 1996.

\bibitem{Venkatasubramanian:2000p7305}
R~Venkatasubramanian.
\newblock Lattice thermal conductivity reduction and phonon localizationlike
  behavior in superlattice structures.
\newblock {\em Phys. Rev. B}, 61(4):3091--3097, 2000.

\bibitem{BorcaTasciuc:2000p15132}
T~Borca-Tasciuc.
\newblock {Thermal conductivity of symmetrically strained Si/Ge superlattices}.
\newblock {\em Superlattices and Microstructures}, 28(3):199--206, 2000.

\bibitem{Chakraborty:2003p15514}
S~Chakraborty, CA~Kleint, A~Heinrich, CM~Schneider, J~Schumann, M~Falke, and
  S~Teichert.
\newblock {Thermal conductivity in strain symmetrized Si/Ge superlattices on Si
  (111)}.
\newblock {\em Applied Physics Letters}, 83:4184, 2003.

\bibitem{Lee:1997p1545}
S~Lee, D~Cahill, and R~Venkatasubramanian.
\newblock {Thermal conductivity of Si--Ge superlattices}.
\newblock {\em Appl. Phys. Lett.}, 70:2957, 1997.

\bibitem{Kasper:1988p15763}
E~Kasper, H~Kibbel, H~Jorke, H~Brugger, E~Friess, and G~Abstreiter.
\newblock {Symmetrically strained Si/Ge superlattices on Si substrates}.
\newblock {\em Physical Review B}, 38(5):3599--3601, Aug 1988.

\bibitem{Kasper:1986p15761}
E~Kasper.
\newblock {Growth and properties of Si/SiGe superlattices}.
\newblock {\em Surface Science}, 174(1-3):630--639, Jan 1986.

\bibitem{Pearsall:1987p15762}
T~Pearsall, J~Bevk, L~Feldman, J~Bonar, J~Mannaerts, and A~Ourmazd.
\newblock Structurally induced optical transitions in {Ge-Si} superlattices.
\newblock {\em Phys. Rev. Lett.}, 58(7):729--732, Feb 1987.

\bibitem{Pearsall:1989p15764}
T~Pearsall, J~Bevk, J~Bean, J~Bonar, J~Mannaerts, and A~Ourmazd.
\newblock Electronic structure of {Ge/Si} monolayer strained-layer
  superlattices.
\newblock {\em Physical Review B}, 39(6):3741--3757, Feb 1989.

\bibitem{Brunner:2002p15513}
K~Brunner.
\newblock {Si/Ge nanostructures}.
\newblock {\em Reports on Progress in Physics}, 65:27, 2002.

\bibitem{Kuan:1991p15765}
T.~S. Kuan and S.~S. Iyer.
\newblock Strain relaxation and ordering in {S}i{G}e layers grown on (100),
  (111), and (110) {S}i surfaces by molecular-beam epitaxy.
\newblock {\em Appl.~Phys.~Lett.}, 59:2242, 1991.

\bibitem{Bass:1990p15515}
JM~Bass and CC~Matthai.
\newblock {Electronic structure of (111) Si/Ge superlattices}.
\newblock {\em J. Phys.: Condens. Matter}, 2:7841, 1990.

\bibitem{Salehpour:1990p15696}
MR~Salehpour and S~Satpathy.
\newblock Comparison of electron bands of hexagonal and cubic diamond.
\newblock {\em Physical Review B}, 41(5):3048, 1990.

\bibitem{Tahini:2012p15819}
H~Tahini, A~Chroneos, R~W Grimes, U~Schwingenschl{\"o}gl, and A~Dimoulas.
\newblock Strain-induced changes to the electronic structure of germanium.
\newblock {\em J. Phys.: Condens. Matter}, 24(19):195802.

\bibitem{Zhang:2009p15487}
Feng Zhang, Vincent Crespi, and Peihong Zhang.
\newblock {Prediction that Uniaxial Tension along ⟨111⟩ Produces a Direct
  Band Gap in Germanium}.
\newblock {\em Phys. Rev. Lett.}, 102(15):156401, 2009.

\bibitem{Rideau:2006p15711}
D~Rideau, M~Feraille, L~Ciampolini, M~Minondo, C~Tavernier, H~Jaouen, and
  A~Ghetti.
\newblock {Strained Si, Ge, and Si$_{1−x}$Ge$_{x}$ alloys modeled with a
  first-principles-optimized full-zone k∙p method}.
\newblock {\em Physical Review B}, 74(19):195208, 2006.

\bibitem{Braunstein:1958p15767}
Rubin Braunstein, Arnold~R. Moore, and Frank Herman.
\newblock Intrinsic optical absorption in germanium-silicon alloys.
\newblock {\em Phys. Rev.}, 109:695--710, 1958.

\bibitem{Ioffe:1960}
A.~F. Ioffe.
\newblock {\em Physics of Semiconductors}.
\newblock Academic, New York, 1960.

\bibitem{Wang:2008p15709}
XW~Wang, H~Lee, YC~Lan, GH~Zhu, G~Joshi, DZ~Wang, J~Yang, AJ~Muto, MY~Tang, and
  J~Klatsky.
\newblock Enhanced thermoelectric figure of merit in nanostructured n-type
  silicon germanium bulk alloy.
\newblock {\em Applied Physics Letters}, 93:193121, 2008.

\bibitem{Pernot:2010p14944}
G~Pernot, M~Stoffel, I~Savic, F~Pezzoli, P~Chen, G~Savelli, A~Jacquot,
  J~Schumann, U~Denker, and I~M{\"o}nch.
\newblock Precise control of thermal conductivity at the nanoscale through
  individual phonon-scattering barriers.
\newblock {\em Nature Materials}, 9:491, 2010.

\bibitem{Jalan:2010p14038}
B~Jalan and Susanne Stemmer.
\newblock {Large Seebeck coefficients and thermoelectric power factor of
  La-doped SrTiO$_3$ thin films}.
\newblock {\em Applied Physics Letters}, 97(4):3, 2010.

\bibitem{Okinaka:2006p15720}
Noriyuki Okinaka and Tomohiro Akiyama.
\newblock {Latent Property of Defect-Controlled Metal Oxide: Nonstoichiometric
  Titanium Oxides as Prospective Material for High-Temperature Thermoelectric
  Conversion}.
\newblock {\em Japanese Journal of Applied Physics}, 45(9):7009--7010, 2006.

\bibitem{Vashaee:2007p15721}
Daryoosh Vashaee and Ali Shakouri.
\newblock Thermionic power generation at high temperatures using {SiGe/Si}
  superlattices.
\newblock {\em Journal of Applied Physics}, 101(5):053719, 2007.

\bibitem{Shakouri:1997p15266}
A~Shakouri and J.E Bowers.
\newblock Heterostructure integrated thermionic coolers.
\newblock {\em Applied Physics Letters}, 71:1234, 1997.

\bibitem{Vashaee:2004p6592}
D~Vashaee and A~Shakouri.
\newblock Improved thermoelectric power factor in metal-based superlattices.
\newblock {\em Phys. Rev. Lett.}, 92(10):106103, 2004.

\bibitem{Sun:2009p15718}
Peijie Sun, Niels Oeschler, Simon Johnsen, Bo~B Iversen, and Frank Steglich.
\newblock {Huge Thermoelectric Power Factor: FeSb$_2$ versus FeAs$_2$ and
  RuSb$_2$}.
\newblock {\em Applied Physics Express}, 2(9):091102, 2009.

\bibitem{Lee:2010p6879}
Joo-Hyoung Lee, Junqiao Wu, and Jeffrey~C Grossman.
\newblock {Enhancing the Thermoelectric Power Factor with Highly Mismatched
  Isoelectronic Doping}.
\newblock {\em Phys. Rev. Lett.}, 104(1):016602, 2010.

\bibitem{Tritt:2004p15755}
G.~S. Nolas and H.~J. Goldsmid.
\newblock {\em Thermal conductivity: theory, properties, and applications},
  chapter 1.4, page 110.
\newblock Kluwer Academic, New York, 2004.

\bibitem{Chaput:2005p1405}
L~Chaput, P~P{\'e}cheur, J~Tobola, and H~Scherrer.
\newblock Transport in doped skutterudites: Ab initio electronic structure
  calculations.
\newblock {\em Phys. Rev. B}, 72, 2005.

\bibitem{FlageLarsen:2011p15689}
E~Flage-Larsen and O~Prytz.
\newblock The lorenz function: Its properties at optimum thermoelectric
  figure-of-merit.
\newblock {\em Applied Physics Letters}, 99(20):202108, 2011.

\end{thebibliography}

\end{document}